\documentclass[lettersize,journal]{IEEEtran}
\usepackage{graphicx}
\usepackage{amssymb}
\usepackage{amsmath}
\usepackage[table]{xcolor}
\DeclareMathOperator{\Tr}{Tr}

\usepackage{algorithmic}
\usepackage{array}
\usepackage[caption=false,font=footnotesize]{subfig}
\usepackage{stfloats}
\usepackage{url}
\usepackage[short]{optidef}
\usepackage{bm}
\usepackage{makecell}
\usepackage{algorithmic}
\usepackage{booktabs}
\usepackage{adjustbox}
\usepackage{comment}
\usepackage{multirow}
\usepackage{cite}
\usepackage[linesnumbered,ruled,vlined]{algorithm2e}
\usepackage{setspace}
\def\BibTeX{{\rm B\kern-.05em{\sc i\kern-.025em b}\kern-.08em
    T\kern-.1667em\lower.7ex\hbox{E}\kern-.125emX}}

\SetKwInput{KwInput}{Input}                
\SetKwInput{KwOutput}{Output}              
\ifCLASSINFOpdf
\else
\fi

\begin{document}
\title{Hybrid Automatic Repeat Request for Downlink Rate-Splitting Multiple Access}

\author{Rafael~Cerna~Loli,~\IEEEmembership{Graduate~Student~Member,~IEEE},
        Onur~Dizdar,~\IEEEmembership{Member,~IEEE},
        Bruno~Clerckx,~\IEEEmembership{Fellow,~IEEE},
        and~Petar~Popovski,~\IEEEmembership{Fellow,~IEEE}
        
\thanks{R. Cerna Loli is supported by a grant provided by the Defence Science and Technology Laboratory (Dstl) Communications and Networks Research Programme.}
\thanks{R. Cerna Loli and B. Clerckx are with the Department of Electrical and Electronic Engineering, Imperial College London, London SW7 2AZ, U.K. B. Clerckx is also with Silicon Austria Labs (SAL), Graz A-8010, Austria (email: rafael.cerna-loli19@imperial.ac.uk; b.clerckx@imperial.ac.uk).}
\thanks{O. Dizdar is with VIAVI Solutions Inc., Stevenage, SG1 2AN, U.K. (email: onur.dizdar@viavisolutions.com).}
\thanks{P. Popovski is with the Department of Electronic Systems, Aalborg University, 9220 Aalborg, Denmark (email: petarp@es.aau.dk).}
}

\maketitle

\begin{abstract}
    This work investigates the design of Hybrid Automatic Repeat Request (HARQ) strategies for downlink Rate-Splitting Multiple Access (RSMA). The existence of private and common stream as well as their conditioning for Successive Interference Cancellation (SIC), gives rise to an expanded set of opportunities for retransmission of failed packets. Specifically, we devise a scheme in which the retransmissions are scheduled through the common stream, which offers a higher success probability. With this, the common stream needs to carry both new and retransmitted bits, which leads to a layered HARQ (L-HARQ) strategy which is capable of trading off throughput and reliability. Simulation results demonstrate that the devised HARQ scheme outperforms RSMA with conventional HARQ, where each retransmission is handled independently through its own stream. It also helps in closing the throughput gap between HARQ and Adaptive Modulation and Coding (AMC) in the high Signal-to-Noise Ratio (SNR) regime while also achieving a decreased Packet Error Rate (PER) and a lower latency.
\end{abstract}

\IEEEpeerreviewmaketitle

\section{Introduction}
Minimal latency and Packet Error Rate (PER) are fundamental features of 5G and beyond mobile communications. For instance, 5G New Radio (NR) mobile communication networks are expected to operate with a latency of 4ms and a PER of $10^{-2} \sim 10^{-4}$ to meet the requirements for enhanced Mobile Broadband (eMBB), and with a latency of 1ms and a PER of $10^{-5}$ to meet the requirements for Ultra Reliable Low Latency Communications (URLLC) \cite{5g_target}. To increase reliability, Hybrid Automatic Repeat Request (HARQ), which merges Forward Error Correction (FEC) coding with Automatic Retransmission Request (ARQ), has been conventionally adopted in wireless networks. With HARQ, a feedback message is sent to the transmitter to request a packet retransmission if a decoding error occurred, or to request a new packet. Although a user can ideally request retransmissions until successfully decoding the packet, in practice only a finite number of retransmissions is allowed before scheduling a new packet so as not to impact the throughput and latency. Therefore, implementing HARQ in applications where only a few retransmissions are allowed is of practical interest.

\begin{figure*}[!t]
		\centering		
        \includegraphics[width=\textwidth]{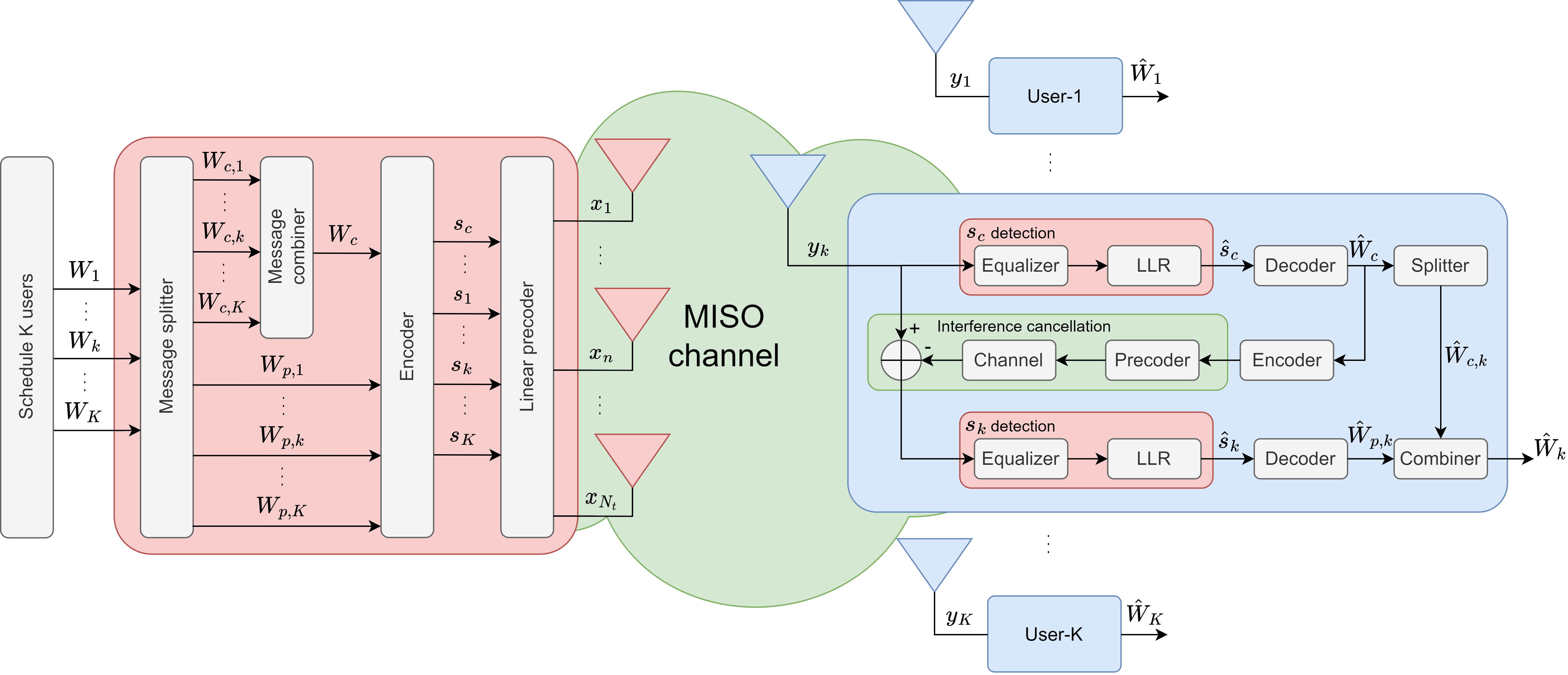}
		\caption{1-Layer RSMA system model \cite{rs_overview_2}.}
		\label{fig:rsma_system_model}
\end{figure*}

In recent years, a new multi-user transmission framework has been introduced, Rate-Splitting Multiple Access (RSMA), which relies on linearly or non-linearly precoded Rate-Splitting (RS) to partially decode the Multi-User Interference (MUI) and partially treat it as noise \cite{eurasip, rs_overview}. As depicted in Fig. \ref{fig:rsma_system_model}, this is achieved by first splitting the user messages and encoding them into common and private streams. Then, all users decode the common stream and, after employing Successive Interference Cancellation (SIC), each of them decodes its intended private stream while considering the private streams of the other users as noise. In this way, RSMA unifies and generalizes other seemingly unrelated transmission schemes, such as Space Division Multiple Access (SDMA), which fully treats MUI as noise, Non-Orthogonal Multiple Access (NOMA), which fully decodes MUI, Orthogonal Multiple Access (OMA), which avoids MUI by transmitting in orthogonal radio resources, and physical-layer multicasting \cite{wcl_2020}. From an information-theoretic perspective, this strategy translates into an increase in the Degrees of Freedom (DoFs) for each user and an increase in the total system Sum-Rate (SR) \cite{joudeh}. It has also been demonstrated that RSMA offers significant benefits in terms of spectral efficiency, reliability, ability to comply with Quality-of-Service (QoS) constraints, robustness against Channel State Information at the Transmitter (CSIT) errors, and receiver complexity reduction, all of which can find novel applications in numerous emerging scenarios for 6G \cite{rs_6g, rs_overview_2,jsac_rs}. However, considering that, to the best of our knowledge, no tailored HARQ scheme for RSMA communications has ever been proposed, designing one is then an important step in the path of standardization for next generation wireless systems.

\subsection{Related works}
HARQ has been applied in multi-user mobile communications since 3G \cite{standard_3g} to the current 5G NR \cite{standard_5g}. In the latter, the 5G Base Station (BS) employs HARQ with Incremental Redundancy (HARQ-IR) to obtain an effective reduction of the transmission rate of the failed packet by means of retransmissions (using the same transmission rate) containing parity/redundancy bits \cite{book_harq_5g} and, hence, an implicit link adaptation is achieved compared to explicitly using the CSIT in every transmission block to adapt the transmission rates. This can improve the system throughput at the expense of incurring in additional delay until the packet is decoded which depends on the number of retransmissions. For this reason, 5G NR systems may also perform user scheduling to prioritize transmitting to users that have requested retransmissions \cite{book_harq_5g_2}. However, this may decrease the individual throughput and increase the latency of other users that would need to wait until resources are available for their own data to be scheduled. Furthermore, recent works have demonstrated that although classical HARQ techniques indeed provide a throughput gain, this is only achieved in the low SNR regime as HARQ schemes prove counterproductive in the high SNR regime due to fixing the transmission rates based on the CSIT in the first transmission block, the first HARQ round, of the HARQ process \cite{amc_harq_foe, l-harq_1}.

Under HARQ transmission, there exists a natural tradeoff between relevant metrics such as throughput, packet reliability, delay, and energy efficiency \cite{harq_survey}. Thus, numerous works have focused on optimizing these HARQ performance metrics in the context of multi-user communications. Regarding throughput improvement, \cite{related_work_1,related_work_2, related_work_3,related_work_4} propose joint rate adaptation (for the first transmission rate in a HARQ process) and user scheduling algorithms to achieve it. Concerning reliability increase, the authors in \cite{related_work_5} derive the expressions for the outage probability and long term average transmission rate of uplink transmission with HARQ in a multi-user Single Input Multiple Output (SIMO) system, and solve the optimal rate selection problem for transmission with HARQ Type-I, HARQ with Chase Combining (HARQ-CC), and HARQ-IR. Also, in \cite{related_work_6}, the latency of multi-user, each with a dedicated channel, single-antenna uplink communications with HARQ-IR is minimized. In order to maximize the energy efficiency under Orthogonal Frequency-Division Multiple Access (OFDMA) transmission (without presence of MUI), the optimal downlink power allocation problem for relay-assisted transmission with HARQ Type-I and average delay constrains per user is solved in \cite{related_work_17}, a dynamic power and bandwidth allocation algorithm considering HARQ Type-I and practical modulation and coding schemes is designed in \cite{related_work_18}, and the effect of the power allocation on the energy-delay tradeoff using HARQ-IR is analyzed in \cite{related_work_19}.

Increasing MUI levels and a lack of radio resources due to spectrum congestion severely limit the performance of 5G NR systems \cite{related_work_20}. Therefore, in recent years NOMA has been increasingly studied to address these challenges due to its capacity to fully decode MUI and serve multiple users using the same radio resources \cite{related_work_21, related_work_22}. Naturally, a trend to combine HARQ with NOMA has also surfaced. In particular, numerous works have focused on characterizing the outage probability of HARQ-enabled NOMA in order to increase the system reliability, design rate selection algorithms, optimize the throughput and power allocation under explicit reliability constraints, or guarantee privacy for secure communications \cite{related_work_7, related_work_8, related_work_9, related_work_10, related_work_11, related_work_12, related_work_23,related_work_24,related_work_25,related_work_26 ,related_work_27,related_work_28,related_work_29,related_work_30}. This is of crucial importance for NOMA as the probability of successful decoding of each user is highly compromised if error propagation occurs along the multiple SIC layers that each user implements. In \cite{related_work_13}, a NOMA-HARQ transmission with blanking is studied in which the theoretical throughput per user is analyzed and the optimization problem of the power and rate allocations is solved to maximize the weighted-sum throughput. Another approach to characterize and improve the performance of NOMA-HARQ schemes is to model the system HARQ processes using a Markov model \cite{related_work_14, related_work_15, related_work_16} to define the different system states and optimize different HARQ metrics. However, all of the previous works are limited in the fact that multi-antenna communications with partial CSIT is not always considered, link adaptation during retransmission rounds is not possible due to the transmission rates being fixed based on the CSIT experienced only during the first HARQ round, and system performance is hardly ever evaluated using practical modulation and coding schemes.
\subsection{Motivation}
Although many studies have been performed for multi-user transmission with HARQ as highlighted in the previous subsection, the performance of RSMA has not been studied in conjunction with HARQ before to the best of the knowledge of the authors. What is more, a customized HARQ scheme for RSMA has never been proposed. Therefore, it is of special interest to address this topic in order to fully define the contribution and importance of RSMA communications in 6G and beyond\cite{rs_overview_2}.

In order to outperform other current multi-user HARQ schemes in a practical deployment, a customized HARQ scheme for RSMA communications must be able to overcome the following challenges:
\begin{itemize}
    \item  Support for multi-antenna communications must be considered: related works only consider single antenna communications in order to simplify the derivation of the HARQ PER expressions. However, it is known that single-antenna schemes are not necessarily efficient in multi-antenna settings \cite{noma_multi_antenna}. 
    \item Fully use the available CSIT in every transmission block in order to select the optimum transmission rates and modulation schemes for each stream in every HARQ round (in every transmission block) in a HARQ process.
    \item Exploit the message splitting and combining operations of RSMA to potentially split and reconstruct the retransmission data if necessary.
    \item Guarantee that the decoding reliability in HARQ retransmission rounds for the private stream is not affected (or not to a high degree) in scenarios where SIC errors occur.
\end{itemize}

\subsection{Main contributions}
The main contributions of this work are summarized as follows:
\begin{itemize}
    \item \textit{First,} to the best of the knowledge of the authors, this is the first paper to propose an advanced HARQ scheme for downlink multi-antenna RSMA communications. The proposed RSMA-HARQ scheme jointly encodes retransmission data and new data packets in a layered HARQ (L-HARQ) approach \cite{l-harq_1, l-harq_2, l-harq_3}. In this way, multiple packets can be transmitted simultaneously in a single stream. Additionally, the transmission rates and modulation schemes can be adaptively adjusted using an Adaptive Modulation and Coding (AMC) algorithm in every transmission block. This contrasts with the HARQ schemes proposed for NOMA and SDMA, which were limited to fixing the transmission rate and modulation schemes during retransmission rounds, and to single antenna transmissions in the former. 
    \item \textit{Second,} by exploiting the message splitting and combining operations of RSMA, it is proposed that the retransmissions should be preferably scheduled through the common stream, as it is innately designed to be decoded with high reliability in order for the SIC process to succeed. By doing this, it is possible for the advanced RSMA-HARQ processes to succeed even when the private stream in the current transmission block is non-decodable. This increased reliability is not achievable with HARQ schemes for NOMA or SDMA, as the retransmissions are always sent through the same stream they were originally scheduled.
    \item \textit{Third,} it is demonstrated through theoretical analysis that the use of the advanced RSMA-HARQ strategy facilitates the otherwise highly intractable calculation of PER expressions of multi-antenna RSMA communications with HARQ, as employing L-HARQ allows for the PER expressions to be decoupled in terms of the Signal-to-Interference-and-Noise Ratio (SINR) experienced across HARQ rounds.
    \item \textit{Fourth,} through link-level simulations, it is demonstrated that the advanced RSMA-HARQ scheme is able to outperform the conventional RSMA transmission without HARQ and the baseline RSMA-HARQ scheme, in which retransmissions are scheduled through the same stream they were originally transmitted through, in terms of throughput, PER, and average latency per bit.
\end{itemize}

\textit{Notation:} Scalars, vectors and matrices are denoted by standard, bold lower and upper case letters, respectively. $\mathbf{I}$ denotes the identity matrix. The Hermitian transpose operator and matrix trace operator are represented by $(.)^H$, and $\text{Tr}(.)$, respectively. The expectation of a random variable is given by $\mathbb{E}\{.\}$. $||.||_2$ is the $l_2$-norm operator, and $\max(.,.)$ is the operator that returns the maximum between its input parameters.  Also, $\mathtt{\sim}$ denotes ``distributed as" and $\mathcal{CN}(0,\sigma^2)$ denotes
the Circularly Symmetric Complex Gaussian (CSCG) distribution with zero mean and variance $\sigma^2$. Finally, we use the superscript $(.)^{(t)}$ to indicate that the parameter is experienced/used in the $t$-th HARQ round (the $t$-th transmission block) of a HARQ process, and the superscript $(.)^{(b,t)}$ is used to denote that the parameter is used in the L-HARQ backtrack decoding operation for a packet that was initially transmitted $t$ HARQ rounds ago.

\section{System Model}
\label{rsma_description}
\subsection{1-Layer RSMA System Model}
 Consider a Base Station (BS), equipped with $N_t$ transmit antennas, that serves $K$ downlink single-antenna users, indexed by the set $\mathcal{K} = \{1,\dots,K\}$, as depicted in Fig. \ref{fig:rsma_system_model}. We consider 1-layer RSMA as described in \cite{eurasip, joudeh, rs_overview, rs_overview_2} which uses a single common stream for all users. In 1-layer RSMA, the message of user-$k$, $W_k$, is split into a common part $W_{c,k}$ and a private part $W_{p,k}$, $\forall k\in\mathcal{K}$. Then, the common parts of all $K$ users $\{W_{c,1},\dots,W_{c,K}\}$ are jointly encoded and modulated into a single common stream $s_c$, while the private parts $\{W_{p,1},\dots,W_{p,K}\}$ are encoded and modulated separately into $K$ private streams $\{s_1,\dots,s_K\}$. The data streams are linearly precoded using the precoder $\mathbf{P} = [\mathbf{p}_c,\mathbf{p}_1,\dots,\mathbf{p}_K] \in \mathbb{C}^{N_t \times (K+1)}$, where $\mathbf{p}_c$ is the common stream precoder and $\mathbf{p}_k$ is the private stream precoder for user-$k$. The transmitted signal $\mathbf{x} \in \mathbb{C}^{N_t \times 1}$, subject to the transmit power constraint $\mathbb{E}\{||\mathbf{x}||^2\}\leq P_t$, is then given by
\begin{equation}
    \mathbf{x} = \mathbf{P}\mathbf{s} = \mathbf{p}_cs_c + \sum_{k=1}^{K}\mathbf{p}_ks_k,
    \label{rsma_transmit_signal}
\end{equation}
where $\mathbf{s} = [s_c,s_1,\dots,s_K]^T \in \mathbb{C}^{(K+1)\times 1}$. It is assumed that $\mathbb{E}\{\mathbf{s}\mathbf{s}^H\}=\mathbf{I}_{(K+1)}$. Hence, $\Tr(\mathbf{P}\mathbf{P}^H)\leq P_t$. At user-$k$, the received signal at the output of the antenna is given by
\begin{equation}
        y_k = \mathbf{h}_k^H\mathbf{P}\mathbf{s} + n_k = \mathbf{h}_k^H\mathbf{p}_cs_c+ \mathbf{h}_k^H\mathbf{p}_k s_k+\overbrace{\sum_{j\neq k,j\in\mathcal{K}}\mathbf{h}_k^H\mathbf{p}_j s_j}^{\text{MUI}} + n_k,
    \label{rsma_eq}
\end{equation}
where $\mathbf{h}_k \in \mathbb{C}^{N_t \times 1}$ is the channel between the transmitter and user-$k$, and $n_k\;\mathtt{\sim}\; \mathcal{CN}(0,\sigma_{n,k}^2)$ is the Additive White Gaussian Noise (AWGN) at user-$k$.

At user-$k$ , $s_c$ is first decoded into $\hat{W}_c$ by treating the interference from the $K$ private streams as noise. Thus, the SINR of decoding $s_c$ is given by
\begin{equation}
    \gamma_{c,k} = \frac{|\mathbf{h}_k^H\mathbf{p}_c|^2}{\sum_{j\in\mathcal{K}}|\mathbf{h}_k^H\mathbf{p}_j|^2+\sigma_{n,k}^2}.
    \label{commonsinr}
\end{equation}
Then, assuming perfect Channel State Information at the Receiver (CSIR), $\hat{W}_c$ is re-encoded, precoded, multiplied by the channel vector and subtracted from $y_k$ using SIC so that the private stream $s_k$ can be decoded into $\hat{W}_{p,k}$ by treating the remaining $K-1$ private streams as noise. The SINR of decoding $s_k$ is then given by
\begin{equation}
    \gamma_{p,k} = \frac{|\mathbf{h}_k^H\mathbf{p}_k|^2}{\sum_{j\neq k,j\in\mathcal{K}}|\mathbf{h}_k^H\mathbf{p}_j|^2+\sigma_{n,k}^2}.
    \label{privatesinr}
\end{equation} 
Finally, user-$k$ extracts $\hat{W}_{c,k}$ from $\hat{W}_c$ and combines it with $\hat{W}_{p,k}$ to reconstruct the message $\hat{W}_k$. Therefore, the achievable rate of the common stream for user-$k$ is  $R_{c,k} = \log_2(1+\gamma_{c,k})$ and the achievable rate of its corresponding private stream is $R_{k} = \log_2(1+\gamma_{k})$. To guarantee that all $K$ users are able to decode the common stream successfully, it must be transmitted at a rate that does not exceed $R_c     = \min\{R_{c,1},\dots,R_{c,K}\}$. User-$k$ is then allocated a portion $C_k$ of the common stream rate $R_c$, and the following relationship is obtained:
\begin{equation}
    R_c=\sum_{k=1}^K C_k\;,\;\forall k \in \mathcal{K}.
\end{equation}

\subsection{Channel State Information Model}
The Channel State Information (CSI) is assumed to follow block-fading. For a given block, the CSI model can be expressed as follows \cite{lina_dpc}:
\begin{equation}
    \mathbf{H}=\hat{\mathbf{H}}+\Tilde{\mathbf{H}},
\end{equation}
where $\mathbf{H}=[\mathbf{h}_1,\dots,\mathbf{h}_K]$ is the real CSI with the entries of $\mathbf{h}_k$ being i.i.d complex Gaussian entries drawn from the distribution $\mathcal{CN}(0,\sigma_k^2), \forall k\in \mathcal{K}$, and $\sigma_k^2$ being the channel amplitude power. Also, $\hat{\mathbf{H}}=[\hat{\mathbf{h}}_1,\dots,\hat{\mathbf{h}}_K]$ is the estimated CSIT with $\hat{\mathbf{h}}_k$ following a Gaussian distribution $\mathcal{CN}(0,\sigma_k^2-\sigma_{e,k}^2), \forall k\in \mathcal{K}$. Finally, $\Tilde{\mathbf{H}}=[\Tilde{\mathbf{h}}_1,\dots,\Tilde{\mathbf{h}}_K]$ represents the CSI estimation error in the CSI estimation/acquisition process, with $\Tilde{\mathbf{h}}_k$ following a Gaussian distribution $\mathcal{CN}(0,\sigma_{e,k}^2), \forall k\in \mathcal{K}$. The parameter $\sigma_{e,k}^2$ is defined as the CSIT error power for user-$k$. The perfect CSIT scenario can then be represented by choosing $\sigma_{e,k}^2 = 0$. In the rest of the paper, we assume that $\sigma_{e,k}^2$ remains constant across transmission blocks.

\section{RSMA-HARQ Transmission}
The design of HARQ in RSMA communications has not been previously studied, as its application is not straightforward  due to the message splitting and combining steps \cite{rs_overview_2}, as shown in Fig. \ref{fig:rsma_system_model}. In this section, the fundamentals of HARQ are first described and two schemes for RSMA-HARQ transmission, baseline and advanced, are introduced next.

\subsection{HARQ Preliminaries}
HARQ schemes aim to increase the reliability of communications by combining retransmissions with FEC coding. In general, they can be classified into three categories \cite{related_work_8}: HARQ Type-I, which uses only the currently received packet (hoping it has the highest SNR across HARQ rounds) for decoding and disregards all other previous erroneous copies; HARQ-CC, which buffers all previous erroneous packets and combines it with the current one using Maximal Ratio Combining (MRC) at the symbol level; and HARQ-IR, which gradually sends parity bits during retransmissions to perform joint decoding of the currently received packet with all other buffered erroneous copies to achieve an effective transmission rate reduction of the original packet. From an information-theoretic perspective, the accumulated mutual information $I^{(T)}$ for each of the three HARQ categories after $T$ HARQ rounds can be expressed as 
\begin{equation}
    I^{(T)} =
    \begin{cases}
        \max\big\{I\big(\gamma^{(t)}\big):t\in[1,T]\big\},&\text{Type-I}\\
        I\big(\sum_{t=1}^T\gamma^{(t)}\big),&\text{CC}\\
        \sum_{t=1}^TI\big(\gamma^{(t)}\big),&\text{IR}
    \end{cases},
\end{equation}
where $I(\gamma)$ denotes the mutual information between channel input and output given the SNR $\gamma$, and $\gamma^{(t)}$ is the decoding SINR in the $t$-th HARQ round. At the end of each HARQ round, the receiver returns a NACK message to the transmitter to request another retransmission if it could not correctly decode the packet, or returns an ACK message to request a new packet after successful decoding. The error event after $T$ HARQ rounds is defined as follows:
\begin{equation}
    \text{ERR}^{(T)}=\big\{I^{(T)}<R^{(1)}\big\},
\end{equation}
where $R^{(1)}=R(\hat{\mathbf{H}}^{(1)})$ is the transmission rate that was chosen in the first HARQ round based on $\hat{\mathbf{H}}^{(1)}$.

The throughput, the long-term average number of decoded bits per channel use, of a HARQ transmission can be employed to characterize its performance and is given by \cite{throughput_ref}
\begin{equation}
        \eta = \lim_{N\rightarrow\infty}\frac{1}{N}\sum_{n=1}^{N}\text{R}[n],
        \label{throughput_eq_1}
    \end{equation}
where $N$ is the number of transmission blocks each spanning $N_s$ symbols, and $\text{R}[n]$, called the instantaneous reward in the $n$-th block, denotes the number of correctly decoded bits in the $n$-th block normalized by $N_s$. During the $T$-th HARQ round of a HARQ process, the instantaneous reward is given by $\text{R}^{(T)}\in \{0, R^{(1)}\}$. Hence, the expected reward $\bar{\text{R}}^{(T)}$, calculated by taking the expectation with respect to the error event $\text{ERR}^{(T)}$, is given by \cite{l-harq_1}
\begin{equation}
    \bar{\text{R}}^{(T)}=R^{(1)}\big(1-\text{PER}(\{\gamma^{(t)}\}_{t=1}^T,R^{(1)})\big),
\end{equation}
where the PER function after $T$ HARQ rounds is given by 
\begin{equation}
    \text{PER}(\{\gamma^{(t)}\}_{t=1}^T,R^{(1)})\triangleq \text{Pr}\big\{\text{ERR}^{(T)}|\{\gamma^{(t)}\}_{t=1}^T,R^{(1)}\big\}.
\end{equation}Thus, it is clear that the throughput performance of HARQ schemes is dependent on knowing the distribution of the combined SINR $\{\gamma^{(t)}\}_{t=1}^T$ and appropriately choosing $R^{(1)}$ according to the constraints on the required number of retransmissions. In this work, we consider HARQ-IR as it is known to outperform both HARQ Type-I and HARQ-CC in terms of PER \cite{related_work_8}. The PER in the $T$-th HARQ round for a HARQ-IR process in the finite block-length regime can be approximated as follows \cite{related_work_15}:
\begin{equation}
\begin{split}
        \text{PER}&(\{\gamma^{(t)}\}_{t=1}^T,R^{(1)}) \approx \\&Q\Bigg(\frac{\sum_{t=1}^T\log_2(1+\gamma^{(t)})-R^{(1)}+\frac{\log_2(TN_s)}{2N_s}}{\sqrt{\frac{\sum_{t=1}^T(1-(1+\gamma^{(t)})^{-2})}{N_s}}\log_2(e)}\Bigg),
\end{split}
\label{harq_ir_per}
\end{equation}
where $Q(.)$ denotes the standard $Q$-function.
\begin{figure*}[!t]
    \centering
    \includegraphics[width=\textwidth]{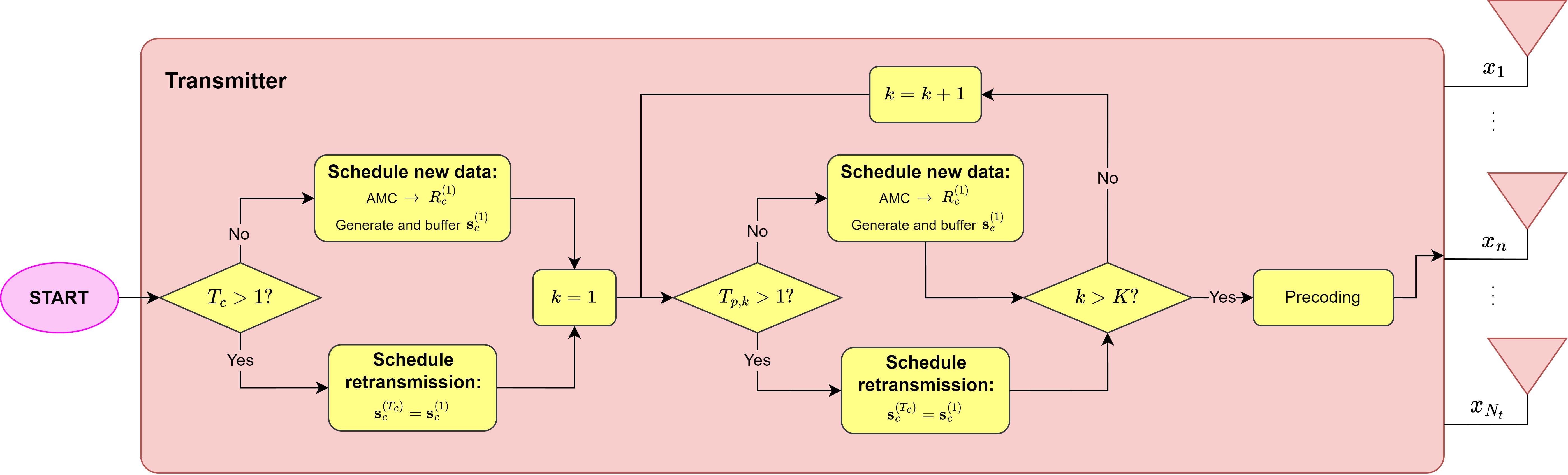}
    \caption{Baseline RSMA-HARQ transmitter operation.}
    \label{fig:baseline_rsma_harq_tx}
\end{figure*}
\subsection{Baseline RSMA-HARQ}
We first define the baseline RSMA-HARQ scheme as the application of HARQ in its simplest and most intuitive form. That is, retransmissions of the common stream are handled through the common stream; and retransmissions of a certain private stream, through the same private stream. The operation of the transmitter and the user receiver of the baseline RSMA-HARQ scheme are depicted in Fig. \ref{fig:baseline_rsma_harq_tx} and Fig. \ref{fig:baseline_rsma_harq_rx}, respectively, and the explanation is as follows:
\begin{figure}[!t]
    \centering
    \includegraphics[width=0.9\columnwidth]{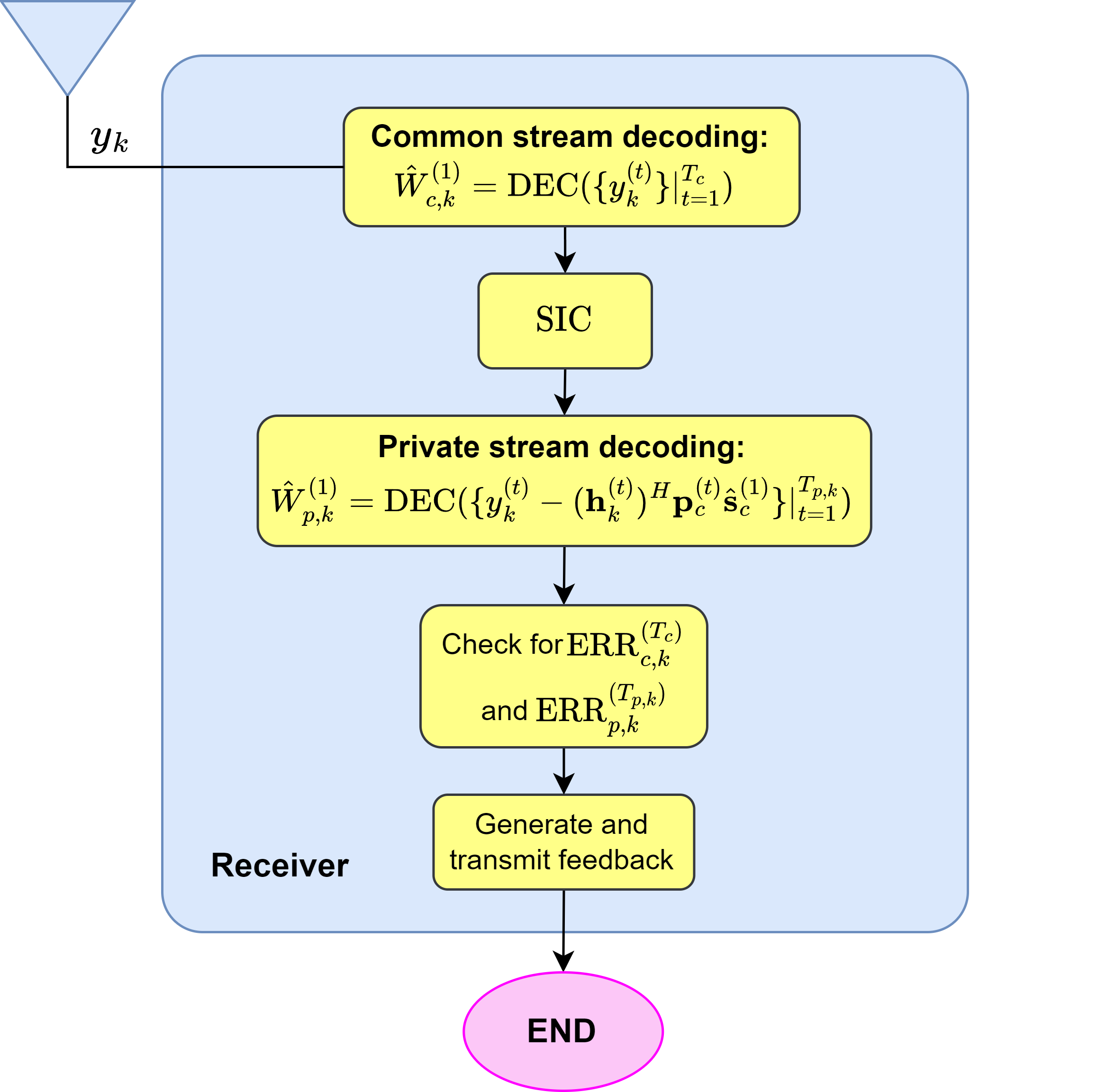}
    \caption{Baseline RSMA-HARQ receiver.}
    \label{fig:baseline_rsma_harq_rx}
\end{figure}
\subsubsection{Transmitter}
At the beginning of each transmission block, the transmitter first checks the index of the current HARQ round $T_c$ of the common stream to schedule new data or a retransmission accordingly. For the scheduling of new data, the transmitter generates $\mathbf{s}_c^{(1)}\in \mathbb{C}^{N_s\times 1}$ as explained in Section \ref{rsma_description}, where $N_s$ is the transmission block length, and stores it in its transmission buffer until the common stream HARQ process finishes. For the scheduling of a retransmission, the transmitter accesses its buffer to update $\mathbf{s}_c^{(T_c)}=\mathbf{s}_c^{(1)}$. The scheduling for each of the private streams is similar. For the $k$-th private stream, the transmitter simply accesses its buffer to update $\mathbf{s}_k^{(T_{p,k})}=\mathbf{s}_k^{(1)}$, where $T_{p,k}$ denotes the current private stream HARQ round, if a retransmission is scheduled. Finally, the transmitter generates the transmit signal $\mathbf{x}$ and transmits it through the channel.
\begin{figure*}[!t]
		\centering
		\includegraphics[width=\textwidth]{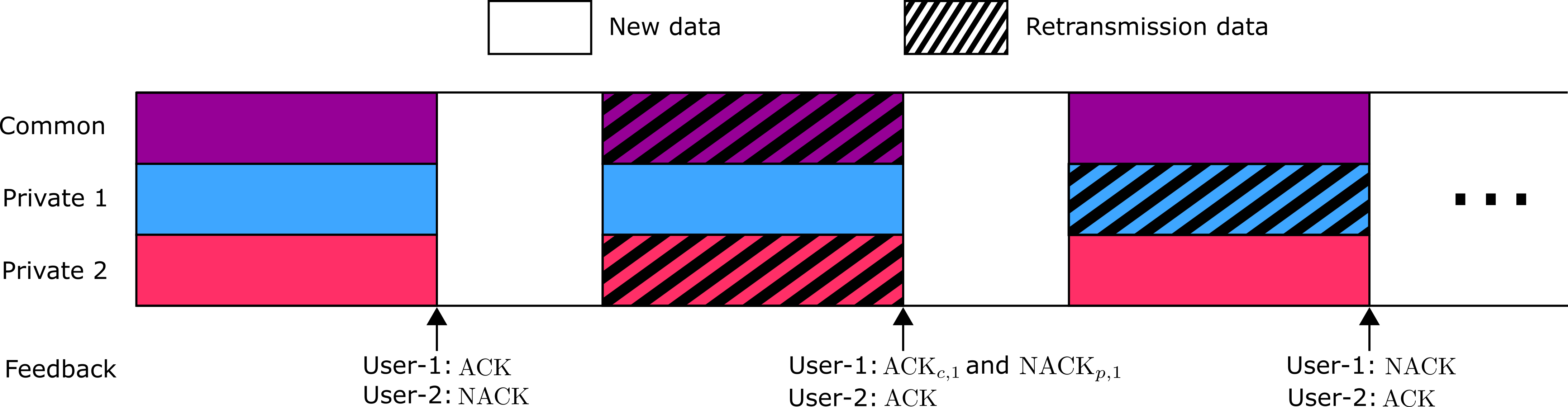}
		\caption{Baseline RSMA-HARQ example with a maximum of 1 retransmission.}
		\label{fig:baseline_rsma_harq}
\end{figure*}
\subsubsection{Receiver}
The receiver at user-$k$ obtains the received signal $y_k$ at the output of its single antenna and attempts to decode the common stream first as $\hat{W}_{c}^{(1)}=\text{DEC}(\{y_k^{(t)}\}|_{t=1}^{T_c})$, where $y_k^{(t)}$ is the received signal buffered in the $t$-th HARQ round and $\text{DEC}(.)$ denotes the decoding function. It then applies SIC to the received signal  and attempts to decode the private stream as $\hat{W}_{p,k}^{(1)}=\text{DEC}(\{y_k^{(t)}-(\mathbf{h}_k^{(t)})^H\mathbf{p}_c^{(t)}\hat{\mathbf{s}}_{c}^{(1)}\}|_{t=1}^{T_{p,k}})$. User-$k$ then checks whether there were errors $\text{ERR}_{c,k}^{(T_c)}$ and $\text{ERR}_{p,k}^{(T_{p,k})}$ in the decoding of the common stream and private stream and generates and transmits the feedback to the transmitter as follows: 
if $\text{ERR}_{c,k}^{(T_c)}$ and $\text{ERR}_{p,k}^{(T_{p,k})}$ occurred, user-$k$ returns a single $\text{NACK}$ feedback to the transmitter. Else, if only $\text{ERR}_{p,k}^{(T_{p,k})}$ occurred, user-$k$ returns a compound feedback $\text{ACK}_{c,k}$ and $\text{NACK}_{p,k}$ to the transmitter. Finally, if no errors occurred, user-$k$ returns a single $\text{ACK}$ feedback to the transmitter. In this way, the total feedback size can be reduced in certain situations, otherwise, user-$k$ would have to transmit feedback for both the common and private stream at the end of every transmission block.
\subsubsection{Expected reward}
Although the distribution of the combined SINR $\{\gamma^{(t)}\}_{t=1}^T$ can be calculated with modest difficulty for single-antenna multi-user communications, this is not straightforward for multi-antenna multi-user communications (and thus for the RSMA framework) due to the presence of multiple fractional random variables in the interference terms of the SINR in equations (\ref{commonsinr}) and (\ref{privatesinr}). Thus, this issue makes the derivation of closed-form expressions for the PER function and the throughput function highly intractable, and ultimately hinders the mathematical analysis of RSMA-HARQ transmission. 

Nevertheless, the performance of different RSMA-HARQ schemes can still be analyzed in a high level manner by defining a general form for the expected reward of the common and private streams and studying their structures. As an example, consider the expected reward of the common stream for user-$k$ in the $T_c$-th HARQ round, which is given by
\begin{equation}
    \bar{\text{C}}_{c,k}^{(T_c)}=C_k^{(1)}\big(1-\text{PER}_{c,k}(\{\gamma_{c,k}^{(t)}\}|_{t=1}^{T_c},R_c^{(1)})\big).
    \label{reward_baseline_common}
\end{equation}
The expected reward of the private stream for user-$k$ in the $T_{p,k}$-th HARQ round can be expressed by 
\begin{equation}         
\begin{split}
    \bar{\text{R}}_{p,k}^{(T_{p,k})}=R_{p,k}^{(1)}&\big(1-\text{PER}_{c,k}(\{\gamma_{c,k}^{(t)}\}|_{t=1}^{T_c},R_c^{(1)})\big)\\
    &\big(1-\text{PER}_{p,k}(\{\gamma_{p,k}^{(t)}\}_{t=1}^{T_{p,k}},R_{p,k}^{(1)})\big).    
\end{split}
 \label{reward_baseline_private}
\end{equation}
It is highlighted that the expected reward of the private stream depends on correctly decoding both the common and private streams, due to the SIC process. Also, the PER terms in (\ref{reward_baseline_common}) and (\ref{reward_baseline_private}) can be approximated using (\ref{harq_ir_per}) as HARQ-IR is applied independently in each stream.
\subsubsection{Example}
To illustrate the operation and challenges of the baseline RSMA-HARQ scheme, consider the example shown in Fig. \ref{fig:baseline_rsma_harq}. At the end of the first transmission block, user-1 correctly decodes the common stream and its private stream and, hence, returns an ACK to the transmitter. However, user-2 is not able to decode the common stream and, consequently, the private stream decoding fails due to error propagation in the SIC process. Thus, user-2 returns a single NACK to the transmitter. In the second transmission block, the transmitter re-schedules the buffered common stream symbols $\mathbf{s}_c^{(1)}$, and also the buffered private stream symbols of user-2 $\mathbf{s}_2^{(1)}$. At the receivers, user-1 directly subtracts $\mathbf{s}_c^{(1)}$, which it already decoded in the previous transmission block, but fails to decode its private stream. Hence, user-1 returns the compound feedback $\text{ACK}_{c,1}$ and $\text{NACK}_{p,1}$ to the transmitter. User-2 applies HARQ to combine the currently received signal $y_2^{(2)}$ with the previous received signal $y_2^{(1)}$ to increase the probability of successful decoding, which is achieved. Therefore, user-2 returns a single ACK as feedback. Finally, in the third transmission block, the transmitter schedules new data in the common stream and private stream of user-2, and the private stream retransmission $\mathbf{s}_1^{(1)}$ for user-1. In this last scenario, user-1 cannot decode the common stream and its private stream, and returns a single NACK, whereas user-2 returns a single ACK as it could decode both streams.   

Three fundamental issues of the baseline RSMA-HARQ are then highlighted. First, it is observed that user-2 was not able to decode the common stream at the end of the first transmission block. Thus, it requests a retransmission which is scheduled again through the common stream. This results in user-1 to not be scheduled new data bits in the common stream in the second transmission block, which ultimately decreases the throughput and increases the latency. Second, the private stream of user-2 could not be decoded at the end of the first transmission block and a retransmission is sent in the second transmission block. However, it is unknown if the error is due to an SIC error or if it would have still happened even with perfect SIC. This results in additional latency and throughput loss in the private stream of user-2. Third, as it happens in general with HARQ schemes, the transmission rates are chosen based only on the CSIT available in the first HARQ round and fixed in subsequent HARQ rounds. Hence, an optimization of the precoder power allocation should be made to compensate this drawback and decrease the PER. However, this is difficult due to not having closed-form expressions or approximations for the PER functions as further discussed in Appendix \ref{app_a}.
\begin{figure*}[!t]
		\centering
		\includegraphics[width=\textwidth]{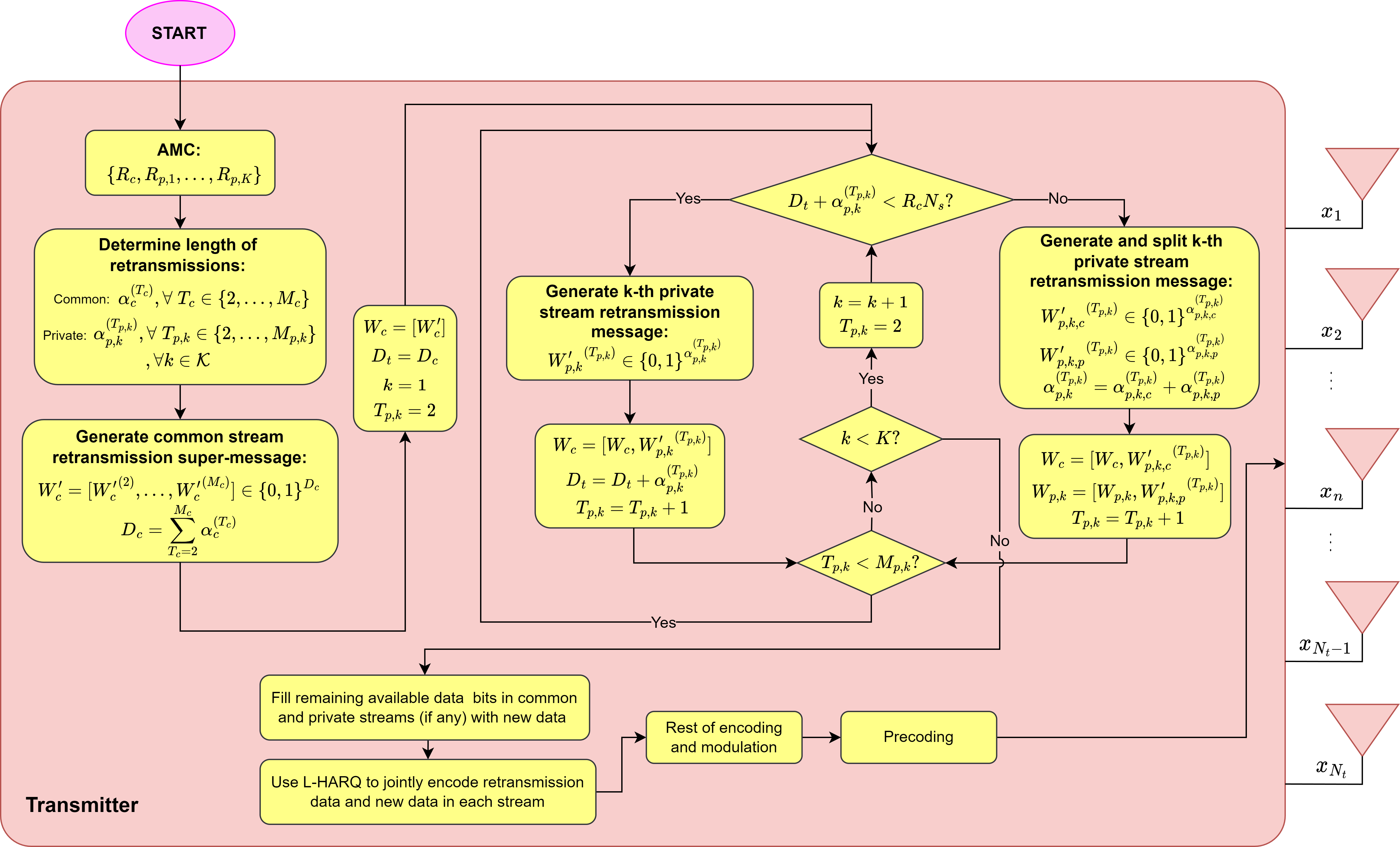}
		\caption{Advanced RSMA-HARQ transmitter operation.}
		\label{fig:advanced_rsma_harq_tx}
\end{figure*}
\subsection{Advanced RSMA-HARQ}
An advanced RSMA-HARQ scheme should then take into account the three shortcomings of the baseline RSMA-HARQ scheme. Namely, it should be able to simultaneously schedule new data and retransmission data through the common stream, mitigate SIC error propagation in order to increase the throughput of the private streams while decreasing their latency, and fully utilize the available CSIT in every transmission block to adapt the transmission rates.

\begin{figure*}[!t]
		\centering
		\includegraphics[width=\textwidth]{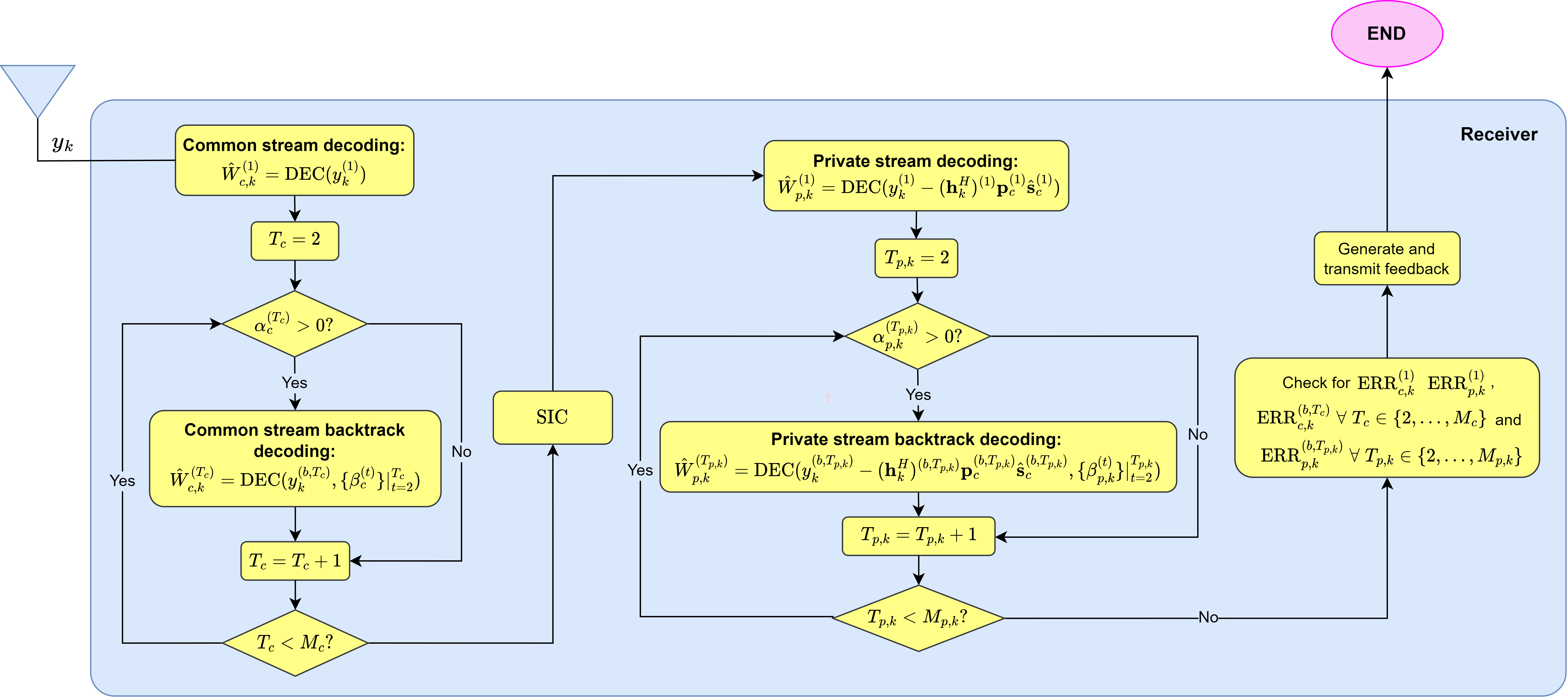}
		\caption{Advanced RSMA-HARQ receiver operation.}
		\label{fig:advanced_rsma_harq_rx}
\end{figure*}

To address these three issues, we propose an advanced RSMA-HARQ scheme that jointly exploits the message splitting and superposed layer structure of RSMA communications, the L-HARQ scheme proposed in \cite{l-harq_1, l-harq_2, l-harq_3}, and AMC. Specifically, all retransmissions are (preferably) scheduled through the common stream, as it should already be designed to be decoded with high reliability in order for SIC to succeed. Each retransmission is also of variable rate: the number of retransmission bits can be adjusted depending on whether reliability or throughput is prioritized. The retransmission data is then jointly re-encoded with new data using the L-HARQ strategy in order to take full advantage of the CSIT in all HARQ rounds. The operation of the transmitter and user receiver of the advanced RSMA-HARQ scheme are depicted in Fig. \ref{fig:advanced_rsma_harq_tx} and Fig. \ref{fig:advanced_rsma_harq_rx}, respectively, and the explanation is as follows:

\subsubsection{Transmitter}
At the beginning of a new transmission block, the transmitter first determines the optimum transmission rates $\{R_c,R_{p,1},\dots,R_{p,K}\}$ using AMC and the current CSIT. Then, it groups the users according to their current HARQ round indices of the common stream into $M_c$ different groups, where $M_c$ is the maximum number of retransmissions for the HARQ processes of the common stream. For group $T_c\;\in \{2,\dots,M_c\}$, the transmitter chooses $\alpha_c^{(T_c)}$, which denotes the number of retransmission bits that will be scheduled in the retransmission for the $T_c$-th common stream HARQ group. This process is similarly repeated for each of the $K$ private streams. For user-$k$, it is possible for the transmitter to simultaneously schedule up to $M_{p,k}$ retransmissions, where $M_{p,k}$ is the maximum number of retransmissions for the HARQ processes of the $k$-th private stream. Thus, for retransmission $T_{p,k}\;\in\{2,\dots,M_{p,k}\}$ of the private stream of user-$k$, the transmitter chooses $\alpha_{p,k}^{(T_{p,k})}$, which denotes the number of retransmission bits that will be scheduled in it.

The retransmissions for the different common stream HARQ groups are first concatenated to obtain the common stream retransmission super-message $W'_c \in \{0,1\}^{D_c}$, where $D_c=\sum_{T_c=2}^{M_c}\alpha_c^{(T_c)}$. Then, the private stream retransmissions are ordered in descending order according to their HARQ round indices and are sequentially concatenated with $W'_c$ until all ${R_c N_s}$ information bits in the common stream are filled. If concatenating a private stream retransmission would result in exceeding the number of available information bits in the common stream, then the private stream retransmission is splitted and $\alpha_{p,k,c}^{(T_{p,k})}$ retransmission bits are scheduled in the common stream to complete scheduling the available information bits, and the remaining $\alpha_{p,k,p}^{(T_{p,k})}$ retransmission bits are scheduled in the respective private stream and re-encoded along with new private stream data using L-HARQ. If there are still information bits available in the common stream after scheduling all private retransmissions, these are used to carry new data in the common stream prioritizing the users that have not been scheduled any data yet. Finally, the transmitter generates the transmit signal $\mathbf{x}$ and transmits it through the channel.
\subsubsection{Receiver}
User-$k$ obtains the received signal $y_k$ at the output of its single antenna and attempts to decode the common stream first as $\hat{W}_{c}^{(1)}=\text{DEC}(y_k^{(1)})$. If any retransmissions for previous common stream data were also scheduled and the common stream was successfully decoded, user-$k$ then performs the backtrack decoding operation of L-HARQ \cite{l-harq_2} $\hat{W}_{c}^{(T_c)}=\text{DEC}(y_k^{(b,T_c)},\{\beta_{c,k}^{(t)}\}|_{t=2}^{T_{c}})$  for each of its scheduled common stream retransmissions, where $y_k^{(b,T_c)}$ is the received signal at user-$k$ that was buffered in the first transmission block of the $T_c$-th common stream HARQ group and $\beta_{c,k}^{(t)}$ denotes the number of successfully extracted common stream retransmission bits from the common stream at user-$k$ during the $t$-th common stream HARQ round. User-$k$ then applies SIC and decodes its private stream as $\hat{W}_{p,k}^{(1)}=\text{DEC}(y_k^{(1)}-(\mathbf{h}_k^H)^{(1)}\mathbf{p}_c^{(1)}\hat{\mathbf{s}}_{c}^{(1)})$. For each of its scheduled private stream retransmissions, user-$k$ recovers the retransmission data, reconstructing it if it was splitted between common and private streams, and performs backtrack decoding $\hat{W}_{p,k}^{(T_{p,k})}=\text{DEC}(y_k^{(b,T_{p,k})}-(\mathbf{h}_k^H)^{(b,T_{p,k})}\mathbf{p}_c^{(b,T_{p,k})}\hat{\mathbf{s}}_{c}^{(b,T_{p,k})},\{\beta_{p,k}^{(t)}\}|_{t=2}^{T_{p,k}})$, where $\beta_{p,k}^{(t)}$ denotes the number of successfully extracted and reconstructed private stream retransmission bits from the common stream and private stream at user-$k$ during the $t$-th private stream HARQ round. 

Then, user-$k$ checks whether there were errors $\text{ERR}_{c,k}^{(1)}$ and $\text{ERR}_{p,k}^{(1)}$ respectively in the decoding of the common stream and private stream, and backtrack decoding errors for its common stream and private stream retransmissions \cite{l-harq_1,l-harq_2}
\begin{equation}
\begin{split}    
    \text{ERR}_{c,k}^{(b,T_c)}&=\text{Pr}\Big\{I(\gamma^{(b,T_c)})<\hat{R}^{(b,T_c)}\Big\},\\    
    \text{ERR}_{p,k}^{(b,T_{p,k})}&=\text{Pr}\Big\{I(\gamma^{(b,T_{p,k})})<\hat{R}^{(b,T_{p,k})}\Big\},
\end{split}
\end{equation}
where $\hat{R}^{(b,T_c)}=R^{(b,T_c)}-\sum_{t=2}^{T_c}\beta_{c,k}^{(t)}/N_s$ and $\hat{R}^{(b,T_{p,k})}=R^{(b,T_{p,k})}-\sum_{t=2}^{T_{p,k}}\beta_{p,k}^{(t)}/N_s$ denote the reduced common and private backtrack decoding rates. Then, user-$k$ generates and transmits the feedback to the transmitter as follows: If all decoding attempts culminated in errors, user-$k$ returns a single $\text{NACK}$ feedback to the transmitter. Else, the transmitter orders the feedback data in the order that decoding occurred and returns the individual results for each of them. In the scenario in which there is a subset at the end of the feedback data with equal results, the transmitter replaces that subset accordingly with a single $\text{ACK}$ or $\text{NACK}$. Finally, if no decoding errors occurred, user-$k$ returns a single $\text{ACK}$ to the transmitter.
\begin{figure*}[!t]
\begin{equation}
        \text{R}_{c,k}^{(T_{c})} = \Bigg[ \sum_{T_b=2}^{T_c} C_k^{(b,T_b)}\big(1-\text{PER}_{c,k}^b(\gamma_{c,k}^{(b,T_b)},\{\beta_{c}^{(t)}\}|_{t=2}^{T_b},R_c^{(b,T_b)})\big)+ C_k^{(1)} \Bigg] \big(1-\text{PER}_{c,k}(\gamma_{c,k}^{(1)},R_c^{(1)})\big),   
\label{reward_advanced_common}
\end{equation}
\end{figure*}
\begin{figure*}[!t]
\begin{equation}
\begin{split}
    \text{R}_{p,k}^{(T_{p,k})} &= \Bigg[\sum_{T_b=2}^{T_{p,k}} R_{p,k}^{(b,T_b)}\Big[\underbrace{\big(\text{PER}_{p,k}(\gamma_{p,k}^{(1)},R_{p,k}^{(1)})\big)
    \big(1-\text{PER}_{p,k}^b(\gamma_{p,k}^{(b,T_b)},\beta_{p,k,c}^{(T_b)},\{\beta_{p,k}^{(t)}\}|_{t=2}^{T_b-1},R_{p,k}^{(b,T_b)})\big)}_{A}\\
    &\;\;\;\;\;+\underbrace{\big(1-\text{PER}_{p,k}(\gamma_{p,k}^{(1)},R_{p,k}^{(1
    )})\big)
    \big(1-\text{PER}_{p,k}^b(\gamma_{p,k}^{(b,T_b)},\{\beta_{p,k}^{(t)}\}|_{t=2}^{T_b},R_{p,k}^{(b,T_b)})\big)}_{B}\Big] \\
    &\;\;\;\;\;+ R_{p,k}^{(1)}\big(1-\text{PER}_{p,k}(\gamma_{p,k}^{(1)},R_{p,k}^{(1)})\big)
    \Bigg]\big(1-\text{PER}_{c,k}(\gamma_{c,k}^{(1)},R_c^{(1)})\big)
    \end{split},
    \label{reward_advanced_private}
\end{equation}
\end{figure*}
\begin{figure*}[!t]
		\centering
		\includegraphics[width=\textwidth]{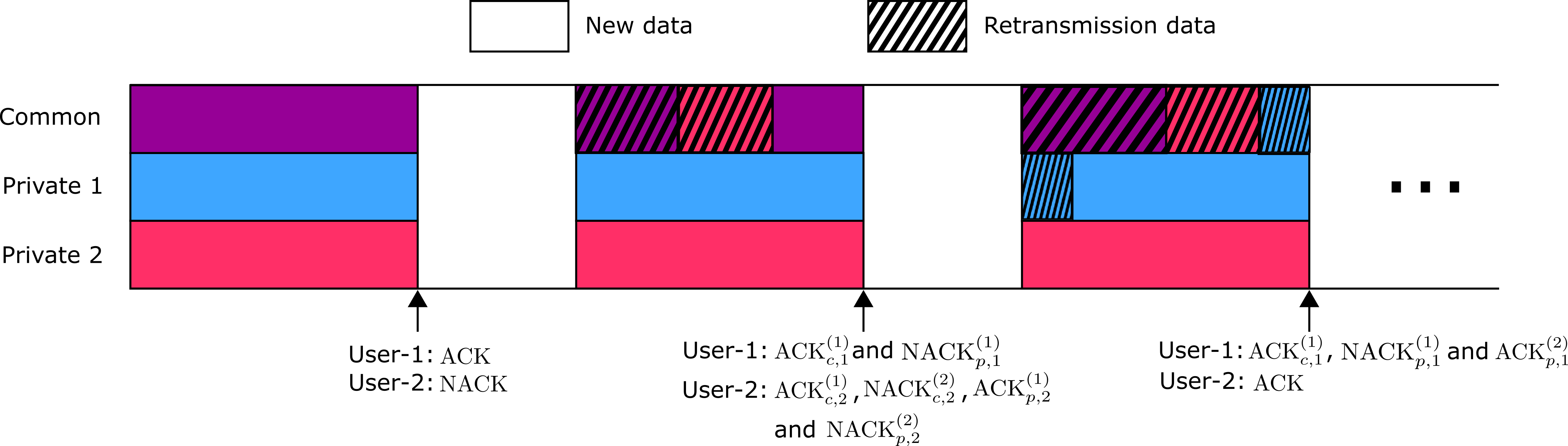}
		\caption{Advanced RSMA-HARQ example with a maximum of 2 retransmissions.}
		\label{fig:proposed_rsma_harq}
\end{figure*}
\subsubsection{Expected reward} 
For user-$k$, consider the expected reward of the common stream during the $T_{c}$-th round of the oldest common stream HARQ process, which is expressed in equation (\ref{reward_advanced_common}) and the expected reward of the private stream in the $T_{p,k}$-th HARQ round of the oldest private stream HARQ process of user-$k$, considering that the retransmission data was splitted between the common and private streams, is expressed in equation (\ref{reward_advanced_private}), where the term $A$ describes the scenario in which the private stream in the $T_{p,k}$-th HARQ round could not be decoded and, hence, backtrack decoding is performed using only the retransmission bits that were scheduled in the common stream. In turn, the term $B$ denotes the scenario in which the private stream could be correctly decoded and backtrack decoding is consequently performed after combining the retransmission bits that were scheduled in the common stream together with the retransmission bits that were scheduled in the private stream. Also, the backtrack PER function of a packet in the $T$-th HARQ round is given by
\begin{equation}
\begin{split}
    \text{PER}^b(\gamma^{(b,T)}&,\{\beta^{(t)}\}|_{t=2}^{T},R^{(b,T)})\triangleq \\&\text{Pr}\big\{\text{ERR}^{(b,T)}|\gamma^{(b,T)},\{\beta^{(t)}\}|_{t=2}^{T},R^{(b,T)}\big\}.
\end{split}
    \label{per_backtrack}
\end{equation}
It can be noticed that the advanced RSMA-HARQ scheme has the potential to achieve higher throughput compared to the baseline RSMA-HARQ due to the simultaneous scheduling of retransmissions and new data. Also, differently from the baseline RSMA-HARQ scheme, all of the decoding steps of the advanced RSMA-HARQ scheme depend only on the SINR experienced during the first HARQ round of the packet to be decoded. Consequently, the PER functions are decoupled in terms of the SINR experienced during the HARQ cycle. This enables the application of AMC in every transmission block to adapt the transmission rates of each stream and also facilitates the calculation of the minimum retransmission length $(\alpha_{c}^{(T_{c})})^*$ and $(\alpha_{p,k}^{(T_{p,k})})^*$ for a target PER $\epsilon$ \cite{l-harq_1}. This is further discussed in Appendix \ref{app_b}.
\subsubsection{Decoding complexity}
A receiver using the advanced RSMA-HARQ scheme may have multiple common stream and private stream packets to decode in a given transmission block and, hence, the decoding complexity may vary across transmission blocks depending on the number of packets to decode. Therefore, assuming that the complexity for a decoding algorithm is $D$, the maximum decoding complexity of the advanced RSMA-HARQ scheme at user-$k$ is $(M_c+M_{p,k})D$, where $M_c$ is the maximum number of retransmissions for the common stream packets and $M_{p,k}$ is the maximum number of retransmissions for the private stream of user-$k$. This is due to the fact that user-$k$ will have at most $M_c$ common stream packets and $M_{p,k}$ private stream packets to decode in its buffer. The decoding algorithm complexity $D$ will depend on the specific channel code used.
\subsubsection{Example}
To illustrate the operation and challenges of the advanced RSMA-HARQ scheme, consider the example shown in Fig. \ref{fig:proposed_rsma_harq}. At the end of the first transmission block, user-1 is able to decode both the common and its private stream whereas user-2 fails decoding both. Hence, user-1 returns a single ACK to the transmitter while user-2 returns a single NACK as feedback.

In the second transmission block, the transmitter determines $\alpha_c^{(2)}$ and $\alpha_{p,2}^{(2)}$ and then re-encodes the retransmission bits for the common and private stream retransmissions for user-2 jointly with new data bits and schedules it trough the common stream. The transmitter then schedules new data bits in both private streams. At the receivers, user-1 is able to decode the common stream but fails to decode its private stream. Thus, it generates the feedback $\text{ACK}_{c,1}^{(1)}$ and $\text{NACK}_{p,1}^{(1)}$ and sends it to the transmitter. In turn, user-2 successfully decodes the common stream and private stream but fails to recover the common data and private data sent in the first transmission block after performing backtrack decoding. Therefore, it generates the feedback $\text{ACK}_{c,2}^{(1)},\text{NACK}_{c,2}^{(2)},\text{ACK}_{p,2}^{(1)}$ and $\text{NACK}_{p,2}^{(2)}$ and sends it back to the transmitter.

In the third transmission block, the transmitter decides to increase the size of the common retransmission bits $\alpha_c^{(3)}$ compared to the retransmission bits sent in the second transmission block in order to increase the probability of decoding at user-2. It then determines $\alpha_{p,2}^{(3)}$ for the private stream retransmission of user-2 and also $\alpha_{p,1}^{(2)}$ for the private stream retransmission of user-1 for the private stream packet that failed in the second transmission block. However, as the combined size of all retransmission bits would exceed the number of information bits that the common stream can carry in the third transmission block, the transmitter decides to split the private stream retransmission data for user-1 between the common and its private stream. Finally, the transmitter jointly encodes all retransmission data, and encodes the private stream data, where it is seen that the partial retransmission bits are jointly encoded with new information bits in the private stream of user-1. At the receivers, user-1 is able to decode the common stream but fails to decode its private stream. After extracting the partial private stream retransmission bits that were scheduled through the common stream, it also recovers the private stream data from the previous transmission block after performing backtrack decoding. Therefore, it returns the feedback $\text{ACK}_{c,1}^{(1)}$, $\text{NACK}_{p,1}^{(1)}$ and $\text{ACK}_{p,1}^{(2)}$. User-2, on the other hand, correctly decodes the current common and private packets, and after performing backtrack decoding after extracting the corresponding retransmission bits, it also recovers the common and private data that were sent in the first transmission block. Thus, it returns a single ACK to the transmitter.

The advantages of the advanced RSMA-HARQ scheme over the baseline RSMA-HARQ scheme are then highlighted. First, due to scheduling the retransmissions of the common and private streams preferably through the common stream, it is possible to schedule new data in all streams in every transmission block. Thus, the users can achieve a higher throughput and lower latency compared to the baseline RSMA-HARQ scheme. Second, scheduling the retransmission through the common stream decreases the private stream PER as the SIC process is bypassed. However, it is worth noticing that the advanced RSMA-HARQ scheme requires more feedback bits in some scenarios and also additional control signalling to inform the users of the bit ranges in which their retransmissions are located in the common and private streams.

\section{Numerical Results}
In this section, the performance of the advanced RSMA-HARQ scheme is evaluated in terms of the throughput, PER, Message Error Rate (MER), and average latency per bit, each averaged over $10^6$ random channel realizations drawn from a Rayleigh distribution. A comparison with the baseline RSMA-HARQ scheme and conventional RSMA without HARQ is also given.

\subsection{Simulation setup}
A transmitter with $N_t=8$ antennas that communicates with $K=4$ users is considered. User channels are generated considering $\sigma_k^2=1, \forall k \in \mathcal{K}$, while the channel estimation error variance for imperfect CSIT is $\sigma_{e,k}^2=P_t^{-\alpha}, \forall k \in \mathcal{K}$, where the CSIT quality scaling factor is $\alpha=0.6$ \cite{joudeh}. Also, the noise variance at the receivers is $\sigma_{n,k}^2=1, \forall k \in \mathcal{K}$. The precoder matrix $\mathbf{P}$ for each of the $10^6$ channel realizations is calculated using the Singular Value Decomposition (SVD) and Maximum Ratio Transmission (MRT) method \cite{lina_dpc}, with $90\%$ of the total power allocated to the common stream while the remaining $10\%$ is equally allocated to each of the private streams. Block fading is assumed with each block spanning $N_s=256$ symbols. An AMC algorithm, which uses Quadrature Amplitude Modulation (QAM) modulation schemes, specifically 4-QAM, 16-QAM, 64-QAM, and 256 QAM, is employed to modulate the data streams, while 5G Low Density Parity Check (LDPC) codes are employed as the channel coding scheme according to \cite{ldpc_5g}. We consider two scenarios with 1 or 2 retransmissions respectively before dropping the packet in order to analyze the performance from a low latency perspective, which is a fundamental characteristic of 5G and beyond communications. Also, the retransmission lengths $\alpha_c, \alpha_{p,k}$ for the common and private streams in the advanced RSMA-HARQ scheme are fixed to $15\%$ of the total number of retransmission bits that would be sent using the baseline RSMA-HARQ in each HARQ round for the sake of simplicity. Finally, we consider that the feedback from each user at the end of each transmission block is always perfect.

The throughput of RSMA, with or without HARQ, is calculated as follows: 
\begin{equation}
\eta^{\text{RSMA}} = \frac{1}{N}\sum_{n=1}^{N}\sum_{k=1}^{K}(\text{C}_k[n]+\text{R}_k[n]),
\label{throughput_eq_rsma}
\end{equation}
where $N$ is the total number of transmission blocks, $\text{C}_k[n]$ denotes the reward in the common stream in the $n$-th block, considering only the successfully decoded bits intended for user-$k$ after discarding all data intended for other users, and $\text{R}_k[n]$ denotes the reward in the private stream for user-$k$ in the $n$-th block.

As the retransmission bits of the private streams may be scheduled through the common stream in the advanced RSMA-HARQ scheme, the average latency per bit is evaluated in order to have a fair comparison with the baseline RSMA-HARQ transmission and RSMA without HARQ. The average latency per bit after $N$ transmission blocks is then defined as follows:
\begin{equation}
    \lambda=\frac{\sum_{n=1}^N \sum_{k=1}^K T_{c,k}[n]B_{c,k}[n]+T_{p,k}[n]B_{p,k}[n]}{\sum_{n=1}^N \sum_{k=1}^K B_{c,k}[n]+B_{p,k}[n]},
\end{equation}
where $B_{c,k}[n]$ is the number of correctly decoded bits of user-$k$ that were first sent in the $n$-th block through the common stream, $B_{p,k}[n]$ is the number of correctly decoded bits of user-$k$ that were first sent in the $n$-th block through its private stream, $T_{c,k}[n]$ is the number of transmission blocks that were required to successfully decode $B_{c,k}[n]$, and $T_{p,k}[n]$ is the number of transmission blocks that were required to successfully decode $B_{p,k}[n]$. A special observation is also made about the average latency per bit of the RSMA without HARQ transmission: since all packets are dropped after 1 unsuccessful decoding attempt, we assume that the failed bits are buffered in a queue and re-scheduled in the next transmission blocks in new packets in order to have a fair comparison with other RSMA-HARQ schemes. Thus, new packets with different rates are transmitted in every block using AMC, but the packets may contain re-scheduled bits with different individual latency.

\begin{figure*}[t!]
\begin{minipage}{.5\linewidth}
\centering
\subfloat[]{\label{throughput_a}\includegraphics[scale=.61]{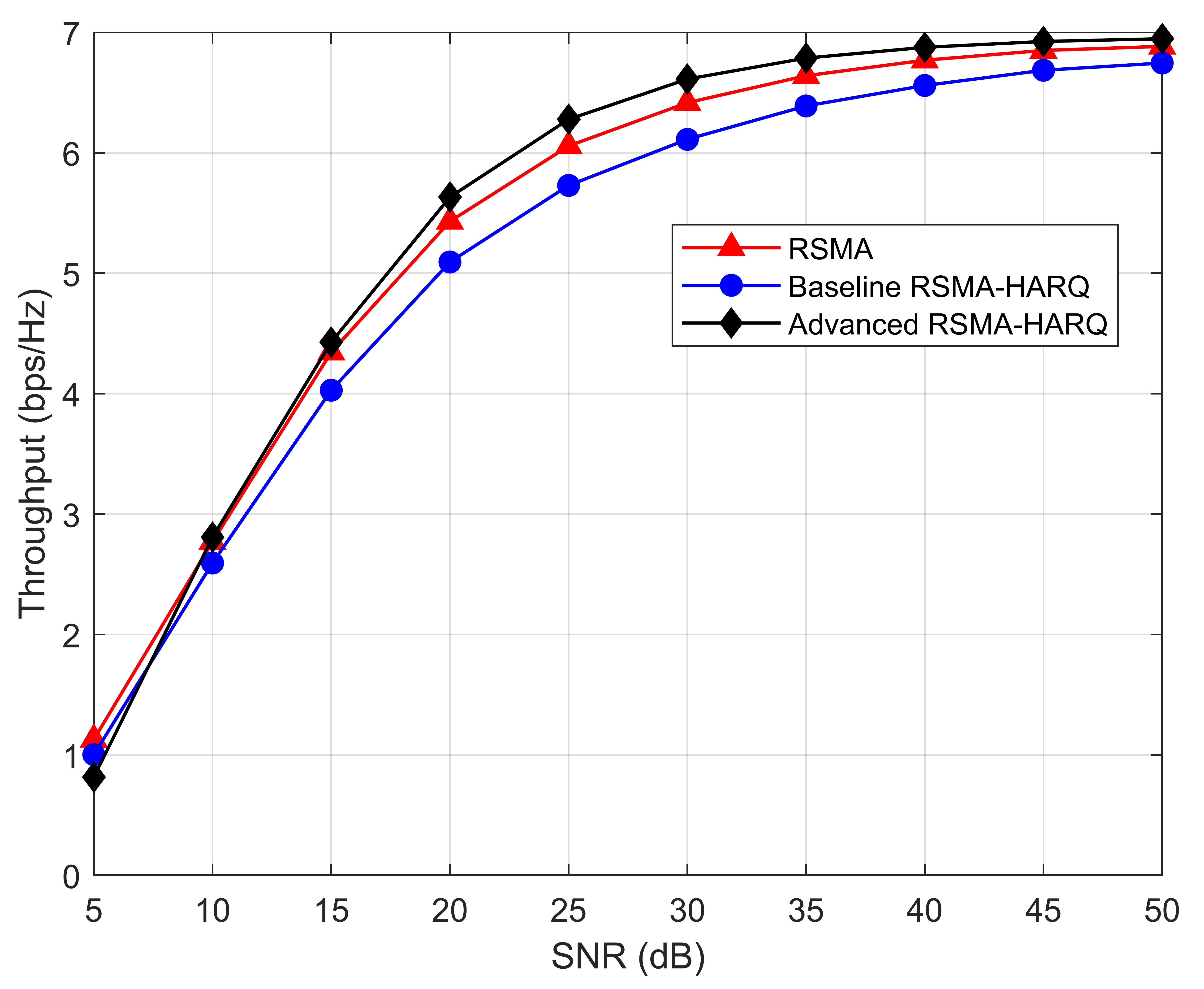}}
\end{minipage}%
\begin{minipage}{.5\linewidth}
\centering
\subfloat[]{\label{throughput_b}\includegraphics[scale=.61]{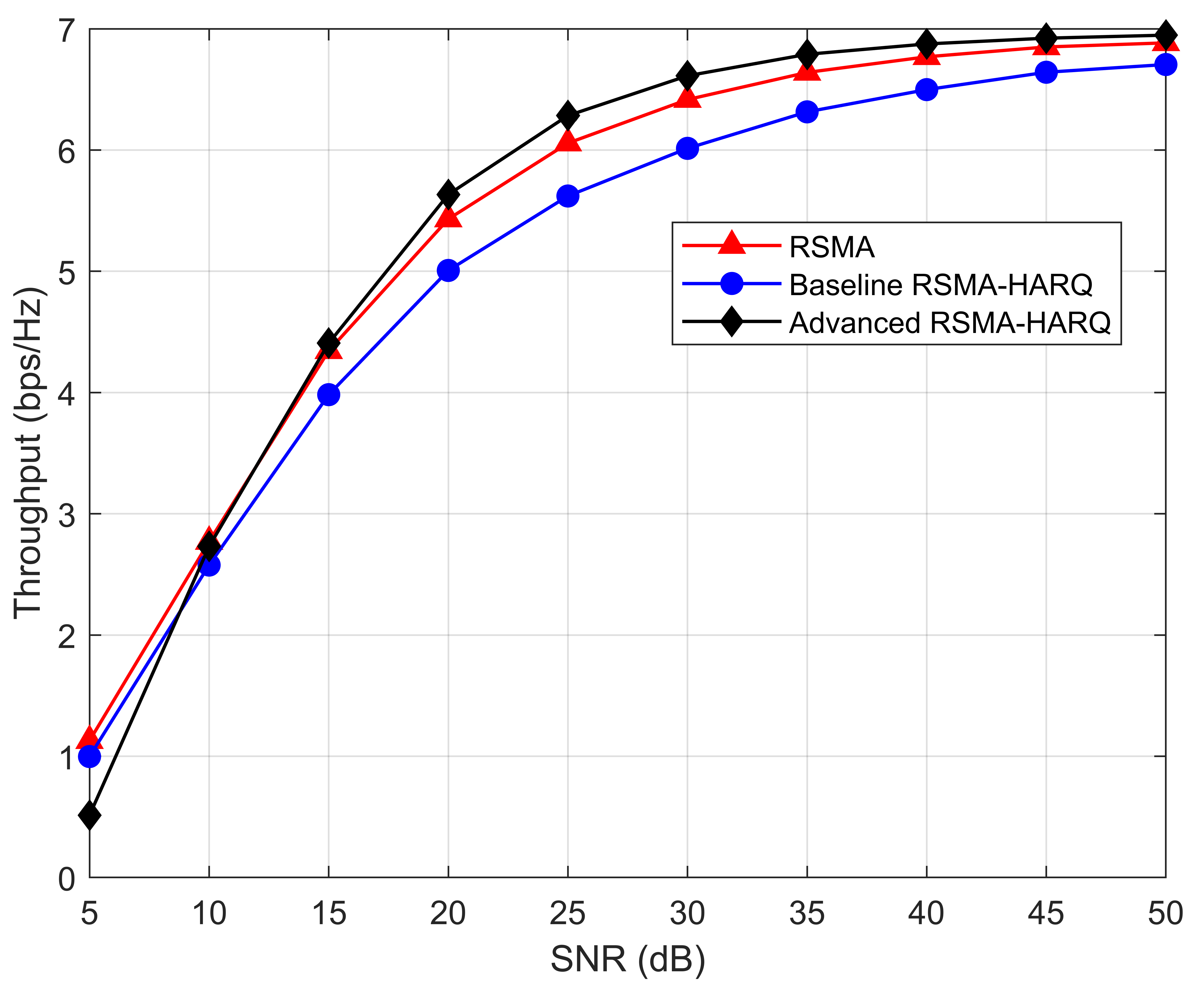}}
\end{minipage}\par\medskip
\caption{Throughput vs. SNR: maximum of (a) 1 retransmission (b) 2 retransmissions.}
\label{fig:throughput}
\end{figure*}
\subsection{Throughput evaluation}
From Fig. \ref{fig:throughput}, it is observed that, in terms of throughput, the advanced RSMA-HARQ scheme outperforms the normal RSMA transmission and the baseline RSMA-HARQ scheme for SNR levels higher than 10 dB for both 1 and 2 retransmissions. As mentioned in \cite{l-harq_2}, the baseline RSMA-HARQ scheme based on conventional application of HARQ underperforms in high SNR levels due to fixing the modulation and transmission rates using only the CSIT available in the first HARQ round, whereas the advanced RSMA-HARQ scheme is able to continuously adapt to changes in the CSIT by using AMC to select the optimum modulation and transmission rates in every block. Additionally, by jointly scheduling the retransmission data and new data through the common stream, the potential achievable throughput of the advanced RSMA-HARQ scheme is larger than normal RSMA transmission with AMC and the baseline RSMA-HARQ scheme, as stated in equations (\ref{reward_advanced_common}) and (\ref{reward_advanced_private}) and also confirmed in Fig. \ref{fig:throughput}.

\begin{figure*}[t!]
\begin{minipage}{.5\linewidth}
\centering
\subfloat[]{\label{per_a}\includegraphics[scale=.61]{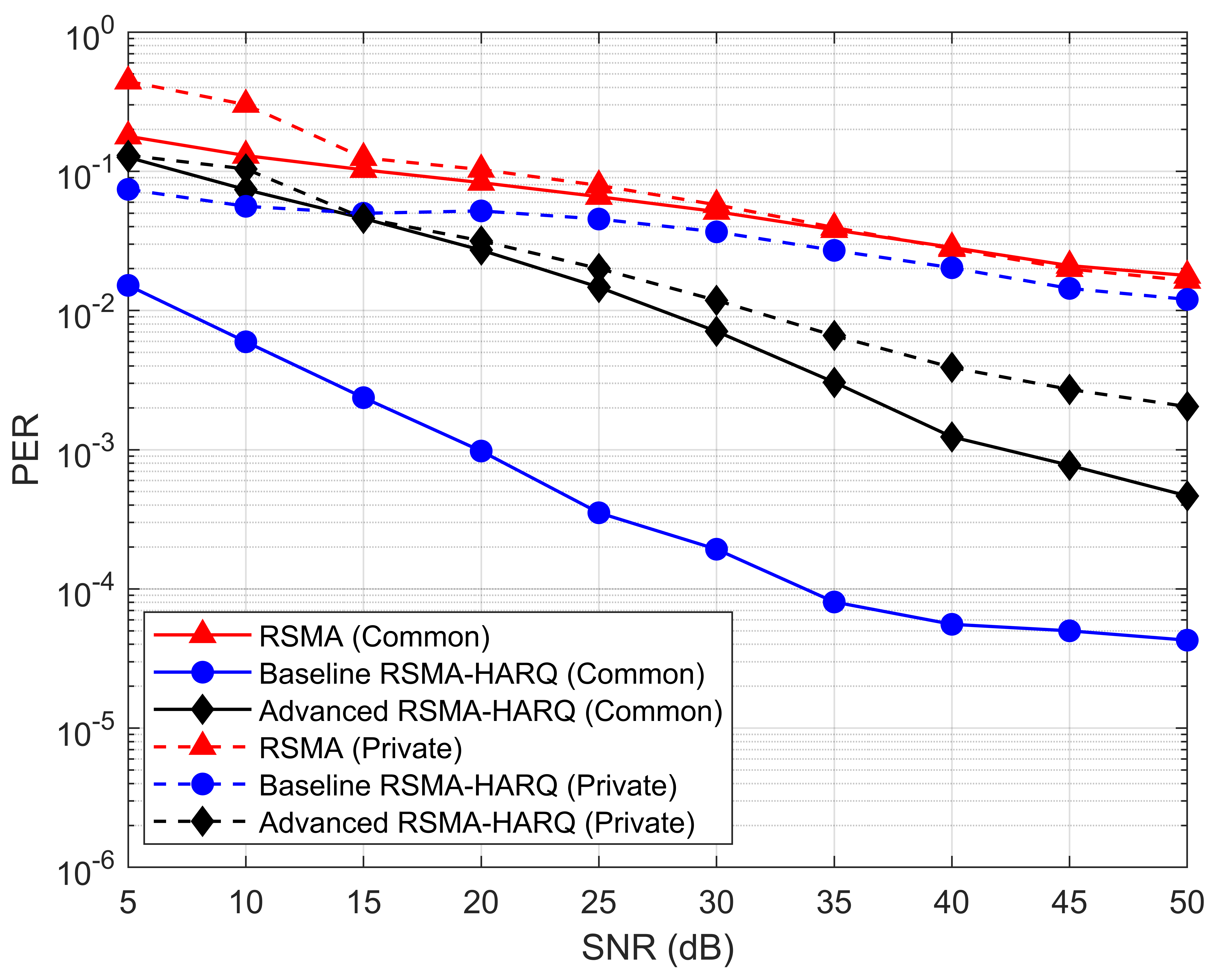}}
\end{minipage}%
\begin{minipage}{.5\linewidth}
\centering
\subfloat[]{\label{per_b}\includegraphics[scale=.61]{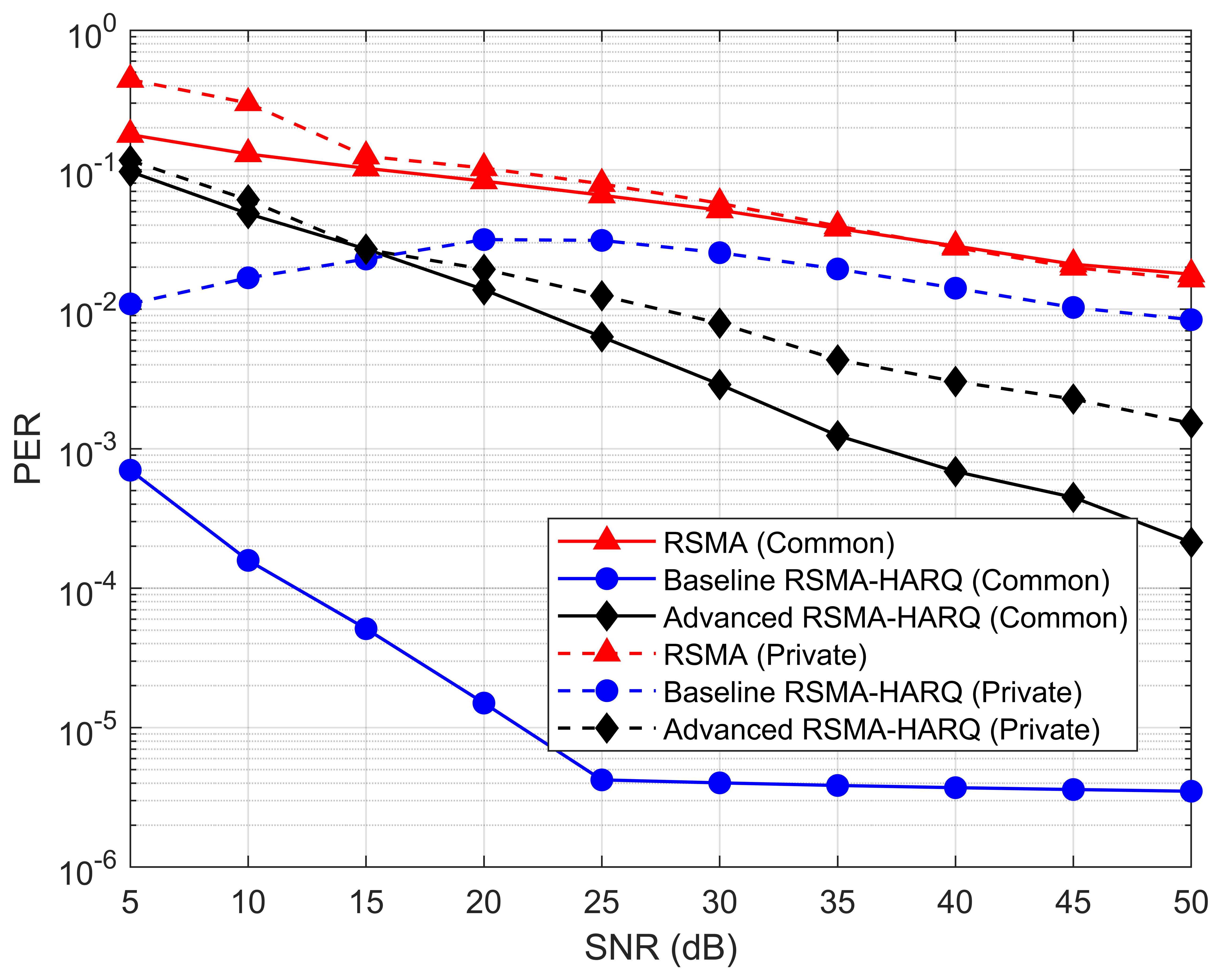}}
\end{minipage}\par\medskip
\caption{PER vs. SNR: maximum of (a) 1 retransmission (b) 2 retransmissions.}
\label{fig:per}
\end{figure*}

In the low SNR regime $(\text{SNR}<10\text{ dB})$, however, the advanced RSMA-HARQ scheme shows the lowest throughput, which is due to the following two reasons: First, due to the larger CSIT error variance in the low SNR regime, the smaller number of retransmission bits scheduled using the advanced RSMA-HARQ scheme are not sufficient to decode a fail packet and, hence, the rewards from failed packets are less likely to be recovered during retransmissions. In contrast, the baseline RSMA-HARQ scheme schedules a much larger number of retransmission bits and is able to recover more rewards from failed packets, as evidenced in the lower PER shown in Fig. \ref{fig:per}. Second, by having to schedule retransmissions more, a lower number of new information bits are scheduled in every transmission block compared to the normal RSMA transmission without HARQ. Thus, although AMC is still used in every transmission block, the rewards recovered by the advanced RSMA-HARQ scheme are lower than those of the normal RSMA transmission, even if the latter achieves a worse PER than the advanced RSMA-HARQ scheme, as evidenced from Fig, \ref{fig:per}. These two reasons suggest that in the low SNR regime, the RSMA transmitter should perform a fallback to the baseline RSMA-HARQ scheme and switch to the advanced RSMA-HARQ scheme as the SNR improves. 

\begin{figure*}[!t]
\begin{minipage}{.5\linewidth}
\centering
\subfloat[]{\label{mer_1}\includegraphics[scale=.61]{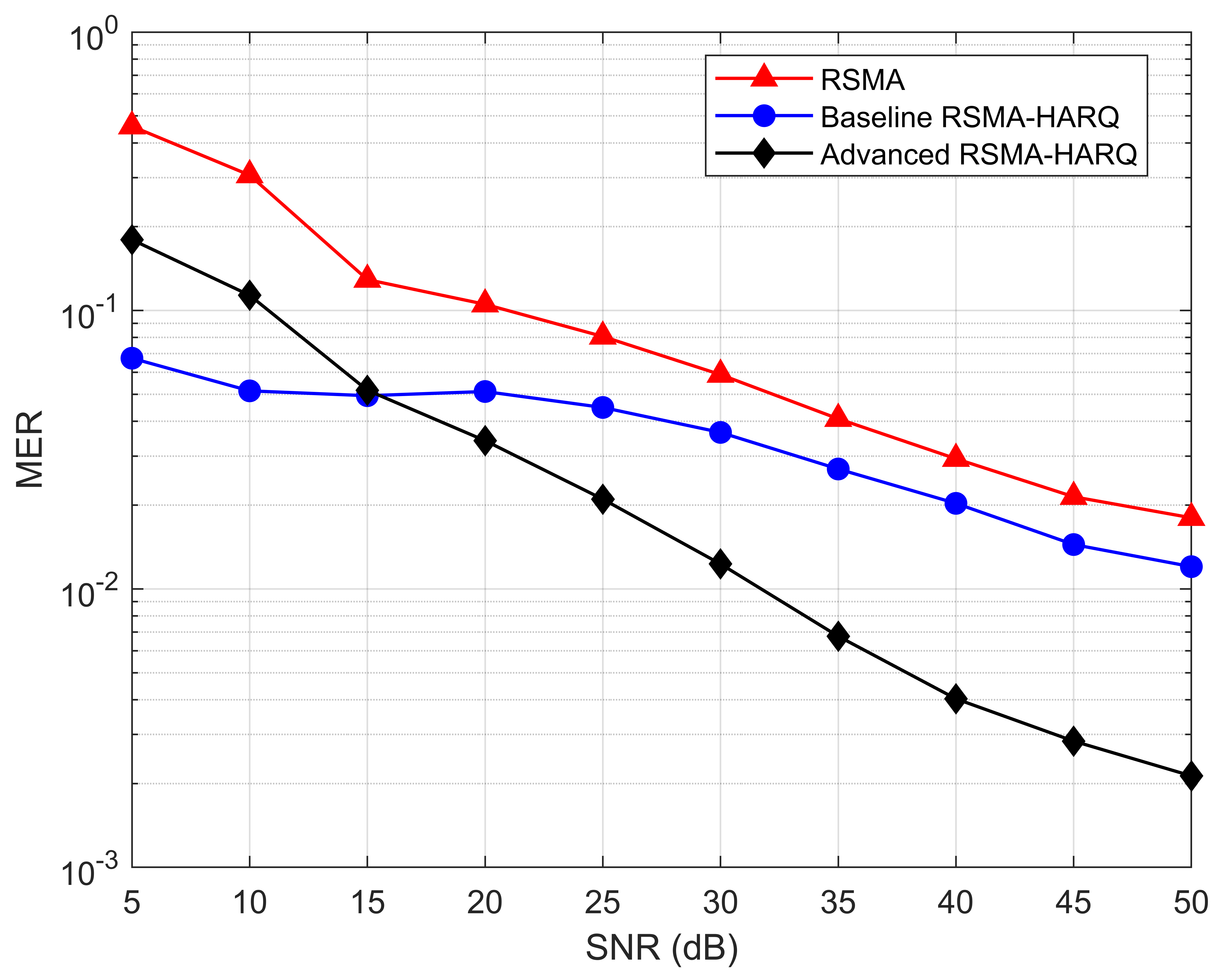}}
\end{minipage}%
\begin{minipage}{.5\linewidth}
\centering
\subfloat[]{\label{mer_2}\includegraphics[scale=.61]{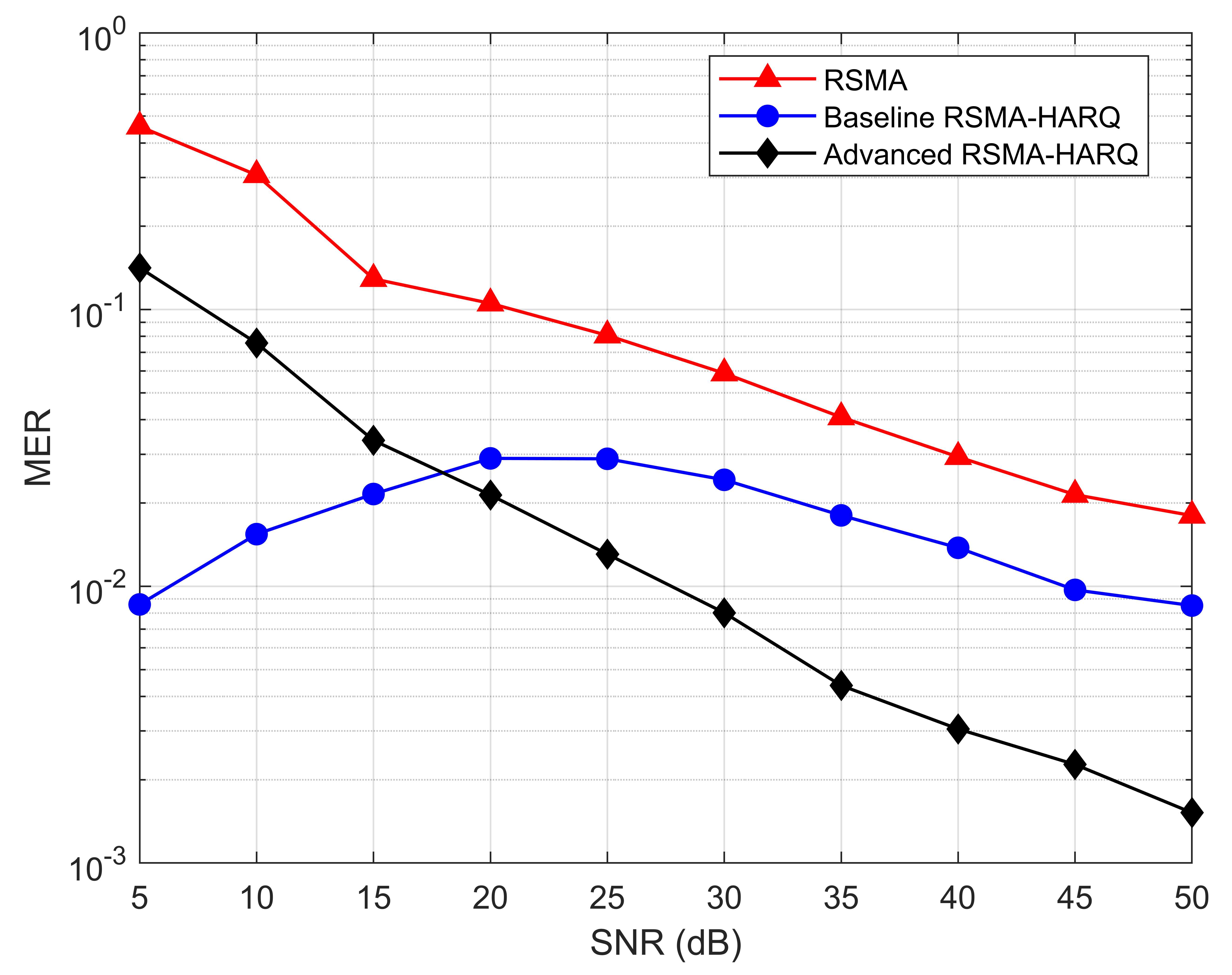}}
\end{minipage}\par\medskip
\caption{MER vs. SNR: maximum of (a) 1 retransmission (b) 2 retransmissions.}
\label{fig:mer_stats}
\end{figure*}
\subsection{PER and MER evaluation}
From Fig. \ref{fig:per}, the main advantage of the advanced RSMA-HARQ scheme is evidenced, as the achieved PER levels for the common and private streams are noticeably lower than those of the normal RSMA transmission with AMC. Compared to the baseline RSMA-HARQ scheme, it can also be observed that the common stream PER is larger while the private stream PER is lower. This demonstrates that the joint application of L-HARQ, AMC, and RSMA transmission to schedule the retransmission data through the common stream is greatly beneficial in terms of PER. Specifically, the unique RSMA stream structure of the common stream allows for the retransmission data to be decoded with high reliability and, consequently, previous decoding errors can be corrected while also freeing space to allocate new data in both the common and private streams. An special observation is then made concerning the trend of the PER curves of the baseline RSMA-HARQ scheme: it is noticed that the private stream PER increases in the low SNR region when considering a maximum of 2 retransmissions. This is due to the fact that the private stream may employ larger modulation schemes than QPSK modulation to transmit with a higher rate as the SNR increases. However, as the baseline RSMA-HARQ fixes the modulation scheme used in retransmissions based on the CSIT of the first HARQ round, retransmissions are more vulnerable to modulation scheme mismatches. This effect is not observed when considering a maximum of 1 retransmission as the incorrect modulation schemes are not fixed for a long duration.

In order to further demonstrate the advantage of the advanced RSMA-HARQ scheme, the Message Error Rate (MER) is plotted in Fig. \ref{fig:mer_stats}. A MER event occurs when the original message cannot be recovered after combining the estimated common and private messages at the receiver. Thus, the MER serves as an ultimate measure of the system error performance. As observed from Fig. \ref{fig:mer_stats}, the baseline RSMA-HARQ scheme achieves the best MER performance for SNR levels below 15 dB. This is due to the larger CSIT error variance in the low SNR region, which increases the likelihood of the AMC algorithm selecting a much larger transmission rate than it is actually supported. Thus, the decoders at the receiver require more retransmission bits to recover failed packets. As the SNR improves, the benefits of the advanced RSMA-HARQ scheme are revealed as the decoders for the common and private streams require less retransmission bits to recover the packets. Additionally, retransmitting the private stream retransmission bits through the common stream increases the reliability of the private stream packets. In contrast, the baseline RSMA-HARQ scheme becomes more susceptible to modulation and transmission rate mismatches by fixing them based only on the CSIT available during the first HARQ round. This is particularly evident from Fig. \ref{mer_2} as it can be seen that the MER increases until the SNR reaches 20 dB. In this region, modulation changes across transmission blocks are frequent. Thus, the MER is more susceptible to using inappropriate modulation and transmission rates.
\begin{figure*}[t!]
\begin{minipage}{.5\linewidth}
\centering
\subfloat[]{\label{latency_a}\includegraphics[scale=.61]{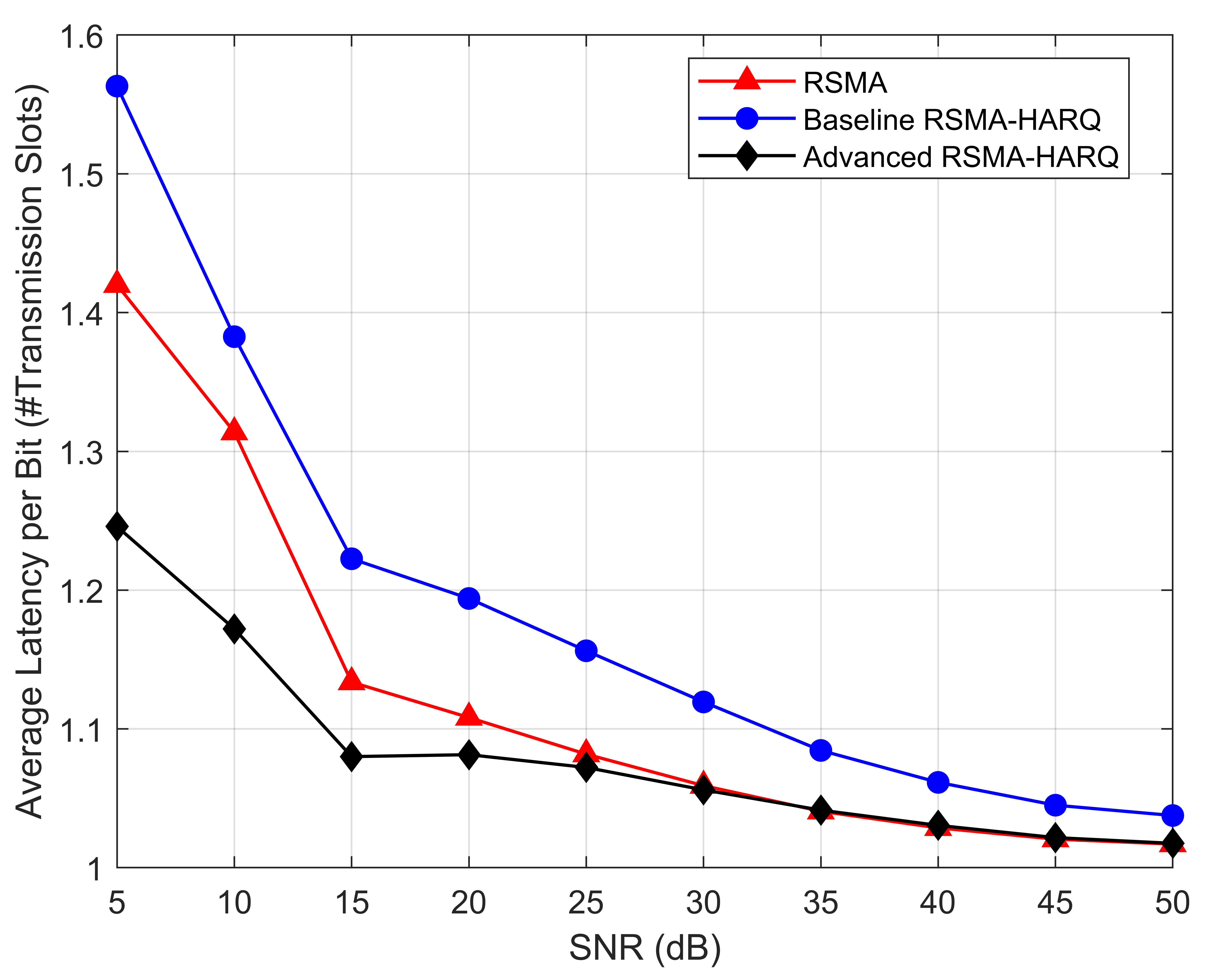}}
\end{minipage}%
\begin{minipage}{.5\linewidth}
\centering
\subfloat[]{\label{latency_b}\includegraphics[scale=.61]{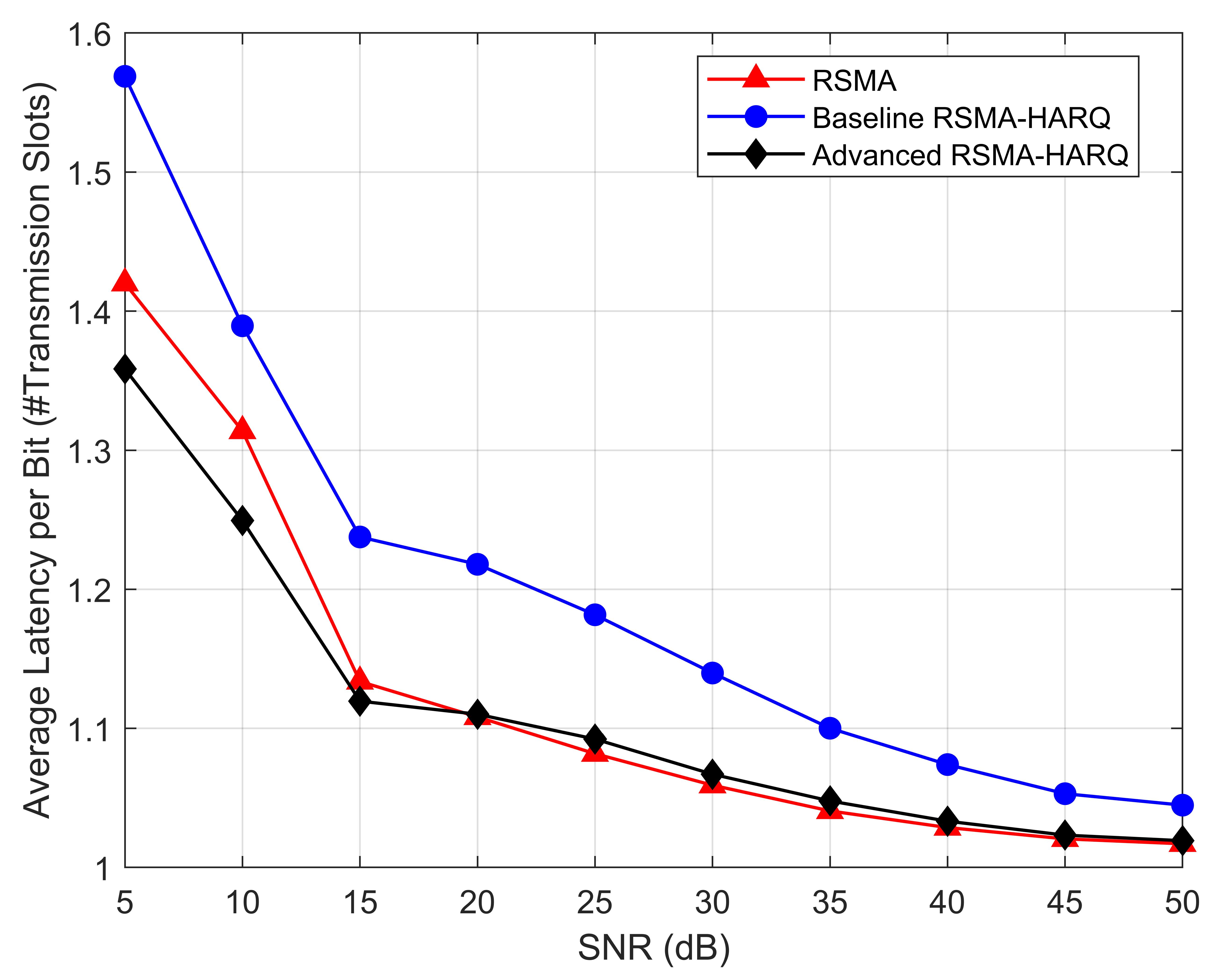}}
\end{minipage}\par\medskip
\caption{Average latency per bit vs. SNR: maximum of (a) 1 retransmission (b) 2 retransmissions.}
\label{fig:latency}
\end{figure*}    
\subsection{Average latency per bit evaluation}
Finally, from Fig. \ref{fig:latency}, it is observed that the advanced RSMA-HARQ scheme outperforms both the baseline RSMA-HARQ scheme and normal RSMA transmission without HARQ in terms of average latency per bit. This is a direct effect of jointly scheduling new data and retransmission data in every transmission block. In contrast, the baseline RSMA-HARQ scheme shows the worst performance as  each user needs to wait for all other users to correctly decode the common message or reach the maximum number of retransmission before being allocated new data. Additionally, private stream retransmissions may not be recovered if the SIC process fails. 

It is also observed that the normal RSMA transmission without HARQ achieves better latency than the baseline RSMA-HARQ scheme due to employing AMC in every transmission block but achieves larger latency than the advanced RSMA-HARQ scheme as every information bit that was not decoded is re-scheduled in the next transmission block, whereas the advanced RSMA-HARQ scheme is able to recover the failed packets by transmitting a smaller number of retransmission bits compared to scheduling all of the failed information bits in the next transmission block. 

\section{Conclusion}
An advanced HARQ scheme for multi-antenna downlink RSMA communications was proposed, which schedules the retransmissions mainly through the common stream in order to maximize packet reliability and throughput, while minimizing the latency. Specifically, retransmission data, either for the common or private stream, is jointly encoded with new data in the common stream using an L-HARQ approach. By appropriately selecting the number of retransmission bits in the retransmission message, the trade-off between minimizing the PER and maximizing the throughput can be adjusted. Numerical results demonstrate that the advanced RSMA-HARQ scheme is able to achieve higher throughput compared to the baseline RSMA-HARQ and normal RSMA transmission without HARQ, while also minimizing the PER and MER, and average latency per bit as the SNR improves.

Future research directions to further optimize the performance of the introduced RSMA-HARQ schemes include a) a precoder power allocation optimization method for the baseline RSMA-HARQ scheme, b) the derivation of the minimum retransmission length for different precoder directions in the advanced RSMA-HARQ scheme, and c) the study of the proposed schemes assuming imperfect feedback in the HARQ process.

\begin{appendix}
\subsection{Average PER of the baseline RSMA-HARQ scheme}
\label{app_a}
The baseline RSMA-HARQ transmitter has access to partial CSIT in every transmission block, albeit it cannot use it to adapt the transmission rates when scheduling retransmissions. Therefore, it may rely on precoder power allocation optimization methods to achieve target PER levels.

For either a common or private packet, the instantaneous PER follows the structure in (\ref{harq_ir_per}). However, this cannot be calculated at the transmitter due to the uncertainty caused by the CSIT error. Therefore, the average PER $\bar{\epsilon}$ is considered instead, which is calculated by taking the expectation of (\ref{harq_ir_per}) over the joint conditional probability density $f_{\{\gamma\}^T|\{\hat{\text{H}}\}^T}(x|\{\mathbf{\hat{H}}^{(t)}\}_{t=1}^T)$ distribution of the combined random SINR $x=\{\gamma^{(t)}\}_{t=1}^T$ given the set of partial CSIT $\{\mathbf{\hat{H}}^{(t)}\}_{t=1}^T$, as follows:
\begin{equation}
    \begin{split}
    \bar{\epsilon}&=\int_0^\infty \text{PER}(\{\gamma^{(t)}\}_{t=1}^T,R^{(1)})f_{\{\gamma\}^T|\{\hat{\text{H}}\}^T}(x|\{\mathbf{\hat{H}}^{(t)}\}_{t=1}^T)dx\\
    &\approx\int_0^\infty Q\Bigg(\frac{\sum_{t=1}^T\log_2(1+\gamma^{(t)})-R^{(1)}+\frac{\log_2(TN_s)}{2N_s}}{\sqrt{\frac{\sum_{t=1}^T(1-(1+\gamma^{(t)})^{-2})}{N_s}}\log_2(e)}\Bigg)\\
    &\quad\quad\quad\quad f_{\{\gamma\}^T|\{\hat{\text{H}}\}^T}(x|\{\mathbf{\hat{H}}^{(t)}\}_{t=1}^T)dx  
    \end{split}.
    \label{expected_per_baseline}
\end{equation}
However, (\ref{expected_per_baseline}) does not have a closed-form expression and the approximation proposed in \cite{makki} cannot be applied as it considers that the received signals in each HARQ round of the HARQ process experience the same SINR. Therefore, its application in optimization methods for the baseline RSMA-HARQ scheme is limited.

\subsection{Average backtrack PER and minimum retransmission length calculation of the advanced RSMA-HARQ scheme}
\label{app_b}
In the common and private expected reward expressions of the advanced RSMA-HARQ scheme in (\ref{reward_advanced_common}) and (\ref{reward_advanced_private}), respectively, it is observed that each of the PER terms depend only on the SINR experienced in the first HARQ round of their corresponding HARQ process. Thus, the overall PER can be optimized by sequentially minimizing the individual PER of each packet in the decoding order followed by the SIC receiver. This decoupling also facilitates the calculation of the minimum retransmission lengths $(\alpha_c^{(T_c)})^*$ and $(\alpha_{p,k}^{(T_{p,k})})^*$.

Consider the general backtrack PER function in (\ref{per_backtrack}) for a packet in the $T$-th HARQ round. In the finite block-length regime, the backtrack PER function can be approximated as follows \cite{backtrack_per, l-harq_1}
\begin{equation}
\begin{split}    
    \text{PER}^b&(\gamma^{(b,T)},\{\beta^{(t)}\}|_{t=2}^{T},R^{(b,T)}) \approx \\&Q\Bigg(\frac{\sqrt{N_s}(\log_2(1+\gamma^{(b,T)})-\hat{R}^{(b,T)})}{\sqrt{(1-(1+\gamma^{(b,T)})^{-2})}\log_2(e)}\Bigg),
\end{split}
\label{per_backtrack_finite}
\end{equation}
where $\hat{R}^{(b,T)}=R^{(b,T)}-\sum_{t=2}^{T}\beta^{(t)}/N_s+\log_2(N_s)/(2N_s)$ is the reduced backtrack decoding rate. Due to partial CSIT, it is not possible to calculate (\ref{per_backtrack_finite}) at the transmitter. Therefore, a more robust approach is to calculate the minimum retransmission length $(\alpha^{(T)})^*$ based on the average backtrack PER $\bar{\epsilon}^{\;b}$, obtained by taking the expectation of (\ref{per_backtrack_finite}) over the conditional probability density distribution $f_{\gamma|\hat{\text{H}}}(x|\mathbf{\hat{H}}^{(b,T)})$ of the random SINR $x=\gamma^{(b,T)}$ given the CSIT $\mathbf{\hat{H}}^{(b,T)}$, as follows:
\begin{equation}
    \begin{split}
    \bar{\epsilon}^{\;b}&=\int_0^\infty \text{PER}^b(\gamma^{(b,T)},\{\beta^{(t)}\}|_{t=2}^{T},R^{(b,T)})f_{\gamma|\hat{\text{H}}}(x|\mathbf{\hat{H}}^{(b,T)})dx\\
    &\approx\int_0^\infty Q\Bigg(\frac{\sqrt{N_s}(\log_2(1+\gamma^{(b,T)})-\hat{R}^{(b,T)})}{\sqrt{(1-(1+\gamma^{(b,T)})^{-2})}\log_2(e)}\Bigg)\\&\quad \quad \quad f_{\gamma|\hat{\text{H}}}(x|\mathbf{\hat{H}}^{(b,T)})dx.  
    \end{split}
    \label{expected_per}
\end{equation}
Based on the work in \cite{makki}, the $Q$-function term in (\ref{expected_per}) can be approximated as follows:
\begin{equation}
    \Xi(\gamma^{(b,T)})=
    \begin{cases}
        1 & \gamma^{(b,T)}\leq\nu\\
        \frac{1}{2}-\lambda(\gamma^{(b,T)}-\xi) & \nu<\gamma^{(b,T)}<\tau\\
        0 & \gamma^{(b,T)}\geq\tau
    \end{cases},
\end{equation}
where 
\begin{align}
    \lambda&=\sqrt{\frac{N_s}{2\pi(2^{2\hat{R}^{(b,T)}}-1)}},\quad\xi=2^{\hat{R}^{(b,T)}}-1,\\
    \nu&=\xi-\frac{1}{2\lambda}\quad\quad\text{and}\quad\quad \tau=\xi+\frac{1}{2\lambda}.
\end{align}
The average backtrack PER $\bar{\epsilon}^{\;b}$ can then be approximated as 
\begin{equation}    \bar{\epsilon}^{\;b}\approx\lambda\int_{\nu}^{\tau}F_{\gamma|\hat{\text{H}}}(x|\mathbf{\hat{H}}^{(b,T)})dx,
\label{avg_backtrack}
\end{equation}
where $F_{\gamma|\hat{\text{H}}}(x|\mathbf{\hat{H}}^{(b,T)})$ is the conditional cumulative distribution function of the random SINR $x=\gamma^{(b,T)}$ given the CSIT $\mathbf{\hat{H}}^{(b,T)}$.

To calculate the minimum retransmission length $(\alpha^{(T)})^*$ for the target PER $\epsilon$, (\ref{avg_backtrack}) can be solved to calculate the value of $(\alpha^{(T)})^*=\beta^{(T)}$ for which the equality $\bar{\epsilon}^{\;b}=\epsilon$ holds. Thus, this only requires the derivation of $F_{\gamma|\hat{\text{H}}}(x|\mathbf{\hat{H}}^{(b,T)})$ which depends on the precoder direction, e.g. Zero-Forcing (ZF), MRT, but is ultimately a simpler task compared to the PER calculation of the baseline RSMA-HARQ scheme, which depends on the joint conditional distribution of multiple SINR realizations and does not have a closed-form expression or approximation.
\end{appendix}

\end{document}